\documentclass[aps,prx,twocolumn,notitlepage,amsfonts,citeautoscript,superscriptaddress,showpacs]{revtex4-1}

\usepackage{natbib}
\usepackage{listings}
\usepackage{color}
\definecolor{dkgreen}{rgb}{0,0.6,0}
\definecolor{gray}{rgb}{0.5,0.5,0.5}
\definecolor{mauve}{rgb}{0.58,0,0.82}

\usepackage{amsmath,amssymb,mathrsfs}
\usepackage{latexsym}
\usepackage{graphicx} 
\usepackage{grffile}
\usepackage{epstopdf}
\usepackage{graphicx,epstopdf,color}
\usepackage{amsfonts}
\usepackage{amsmath,amssymb,mathrsfs}
\usepackage{hyperref}
\usepackage[dvipsnames]{xcolor}
\usepackage{url}
\usepackage{soul}
\usepackage{cancel}

\usepackage{cleveref}
\crefname{equation}{Eq.}{Eqs.}
\Crefname{equation}{Equation}{Equations}
\crefname{figure}{Fig.}{Figs.}
\Crefname{figure}{Figure}{Figures}
\crefname{section}{Sec.}{Secs.}
\Crefname{section}{Section}{Sections}
\crefname{appendix}{Appendix}{Apps.}
\Crefname{appendix}{Appendix}{Apps.}
\crefname{paragraph}{Sec.}{Secs.}
\crefname{table}{Table}{Tables}

\usepackage{bm}
\newcommand{\textalert}[1]{}

\newcommand{\ket}[1]{\left|#1\right\rangle}
\newcommand{\bra}[1]{\left\langle#1\right|}

\newcommand{\om}{\omega} 
\newcommand{\wdr}{\om_\text{d}} 

\def\ie{i.e.\ }
\def\eg{e.g.\ }

\usepackage{xr}

\graphicspath{{nb_fig/}}

\usepackage[caption=false]{subfig}
\usepackage{epstopdf}

\newcommand{\ECa}{E_{\text{C}a}}
\newcommand{\ECb}{E_{\text{C}b}}
\newcommand{\ECc}{E_{\text{C}c}}

\newcommand{\pha}{\hat{\varphi}_a}

\newcommand{\bnj}{\hat{\bm{n}}_j}
\newcommand{\bnk}{\hat{\bm{n}}_k}
\newcommand{\bna}{\hat{\bm{n}}_a}
\newcommand{\bnb}{\hat{\bm{n}}_b}
\newcommand{\bnc}{\hat{\bm{n}}_c}
\newcommand{\bphj}{\hat{\bm{\varphi}}_j}

\newcommand{\bpha}{\hat{\bm{\varphi}}_a}

\newcommand{\bphc}{\hat{\bm{\varphi}}_c}

\newcommand{\phx}{\varphi_\text{ext}}
\newcommand{\bphx}{\overline{\varphi}_\text{ext}}

\newcommand{\EJa}{E_{\text{J}a}}
\newcommand{\EJb}{E_{\text{J}b}}
\newcommand{\EJc}{E_{\text{J}c}}
\newcommand{\ECab}{E_{\text{C}ab}}
\newcommand{\ECbc}{E_{\text{C}bc}}
\newcommand{\ECca}{E_{\text{C}ca}}

\newcommand{\ZZ}{\chi_{ab}}
\newcommand{\J}{J_{ab}}

\newcommand{\hH}{\hat{H}}
\newcommand{\hI}{\hat{H}_\text{I}}
\newcommand{\hG}{\hat{G}}

\newcommand{\wa}{\omega_a}
\newcommand{\wb}{\omega_b}
\newcommand{\wc}{\omega_c}

\newcommand{\uaa}{u_{aa}}
\newcommand{\uab}{u_{ab}}
\newcommand{\uac}{u_{ac}}
\newcommand{\uaj}{u_{aj}}
\newcommand{\uba}{u_{ba}}
\newcommand{\ubb}{u_{bb}}
\newcommand{\ubc}{u_{bc}}
\newcommand{\ubj}{u_{bj}}
\newcommand{\uca}{u_{ca}}
\newcommand{\ucb}{u_{cb}}
\newcommand{\ucc}{u_{cc}}
\newcommand{\ucj}{u_{cj}}

\newcommand{\Wa}{\bm{\omega}_a}
\newcommand{\Wb}{\bm{\omega}_b}
\newcommand{\Wc}{\bm{\omega}_c}

\newcommand{\ha}{\hat{a}}
\newcommand{\hb}{\hat{b}}
\newcommand{\hc}{\hat{c}}

\newcommand{\hba}{\hat{\bm{a}}}
\newcommand{\hbb}{\hat{\bm{b}}}
\newcommand{\hbc}{\hat{\bm{c}}}

\newcommand{\ala}{\alpha_a}
\newcommand{\alb}{\alpha_b}
\newcommand{\alc}{\alpha_c}

\newcommand{\bala}{\bm{\alpha}_a}
\newcommand{\balb}{\bm{\alpha}_b}
\newcommand{\balc}{\bm{\alpha}_c}
\newcommand{\balj}{\bm{\alpha}_j}

\newcommand{\bHI}[1]{\overline{\hH}_I^{(#1)}}
\newcommand{\tHI}[1]{\widetilde{\hH}_I^{(#1)}}
\newcommand{\hGI}{\hat{G}_\text{I}}

\makeatletter
\def\@fnsymbol#1{\ensuremath{\ifcase#1\or * \or \mathsection\or \mathparagraph\or \|\or **\or \else\@ctrerr\fi}}
\makeatother

\begin{document}
\title{Accurate methods for the analysis of strong-drive effects in parametric gates}

\author{Alexandru Petrescu}
\thanks{alexandru.petrescu@usherbrooke.ca}
\affiliation{Institut Quantique and D\'epartement de Physique, Universit\'e de Sherbrooke, Sherbrooke, Qu\'ebec, J1K 2R1, Canada}

\author{Camille Le Calonnec}
\affiliation{Institut Quantique and D\'epartement de Physique, Universit\'e de Sherbrooke, Sherbrooke, Qu\'ebec, J1K 2R1, Canada}

\author{Catherine Leroux}
\affiliation{Institut Quantique and D\'epartement de Physique, Universit\'e de Sherbrooke, Sherbrooke, Qu\'ebec, J1K 2R1, Canada}

\author{Agustin Di Paolo}
\thanks{Current affiliation: Research Laboratory of Electronics, Massachusetts Institute of Technology, Cambridge, MA 02139, USA}
\affiliation{Institut Quantique and D\'epartement de Physique, Universit\'e de Sherbrooke, Sherbrooke, Qu\'ebec, J1K 2R1, Canada}

\author{Pranav Mundada}
\affiliation{Department of Electrical Engineering, Princeton University, Princeton, NJ 08544, USA}

\author{Sara Sussman}
\affiliation{Department of Physics, Princeton University, Princeton, NJ 08544, USA}

\author{Andrei Vrajitoarea}
\affiliation{Department of Electrical Engineering, Princeton University, Princeton, NJ 08544, USA}

\author{Andrew A. Houck}
\affiliation{Department of Electrical Engineering, Princeton University, Princeton, NJ 08544, USA}

\author{Alexandre Blais}
\affiliation{Institut Quantique and D\'epartement de Physique, Universit\'e de Sherbrooke, Sherbrooke, Qu\'ebec, J1K 2R1, Canada}
\affiliation{Canadian  Institute  for  Advanced  Research,  Toronto,  Ontario  M5G  1M1,  Canada}

\date{\today}
\begin{abstract}
  The ability to perform fast, high-fidelity entangling gates is an important requirement for a viable quantum processor. In practice, achieving fast gates often comes with the penalty of strong-drive effects that are not captured by the rotating-wave approximation. These effects can be analyzed in simulations of the gate protocol, but those are computationally costly and often hide the physics at play. Here, we show how to efficiently extract gate parameters by directly solving a Floquet eigenproblem using exact numerics and a perturbative analytical approach. As an example application of this toolkit, we study the space of parametric gates generated between two fixed-frequency transmon qubits connected by a parametrically driven coupler. Our analytical treatment, based on time-dependent Schrieffer-Wolff perturbation theory, yields closed-form expressions for gate frequencies and spurious interactions, and is valid for strong drives. From these calculations, we identify optimal regimes of operation for different types of gates including $i$SWAP, controlled-Z, and CNOT. These analytical results are supplemented by numerical Floquet computations from which we directly extract drive-dependent gate parameters. This approach has a considerable computational advantage over full simulations of time evolutions.  More generally, our combined analytical and numerical strategy allows us to characterize two-qubit gates involving parametrically driven interactions, and can be applied to gate optimization and cross-talk mitigation such as the cancellation of unwanted $ZZ$ interactions in multi-qubit architectures.
\end{abstract}
\maketitle

\section{Introduction}
With considerable advances in state preparation, gate operation, measurement fidelity, and coherence time, superconducting qubits have become one of the leading platforms for quantum information processing \cite{kjaergaard_et_al_2019,krantz_et_al_2019,blais_et_al_2021}. Systems consisting of up to a few dozen qubits have been recently deployed by a number of research groups \cite{versluis_et_al_2017,arute_et_al_2019,Corcoles_2020}. As these architectures are scaled up, an important challenge is to engineer two-qubit interactions to realize gates that are fast enough compared to the decoherence times of the qubits, while at the same time obtaining operation fidelities that are sufficiently high to satisfy a threshold for quantum error correction \cite{gottesman1997stabilizer, aharonov_ben-or_2008}. To realize fast and high-fidelity two-qubit gates, precise modeling of the dynamics of small multi-qubit systems is necessary, but becomes computationally difficult as the number of degrees of freedom increases. Moreover, to achieve fast gates, drives that are strong in the sense of the rotating-wave approximation (RWA) are necessary, in which case beyond-RWA corrections become important.

\begin{figure}[t!]
  \includegraphics{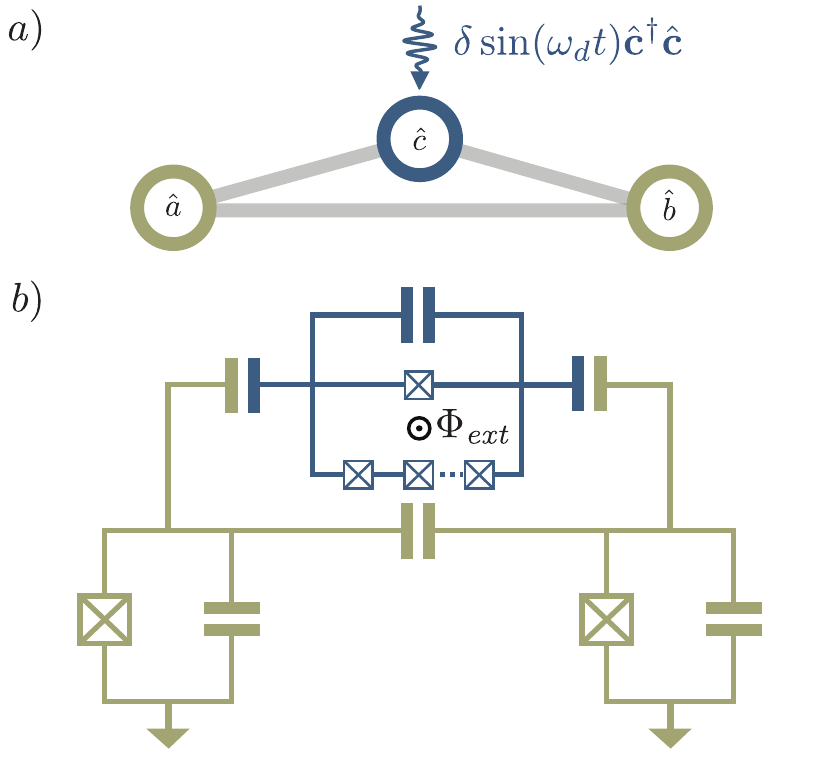}
    \caption{a) Graph representation of the model. The three bare modes have mutual capacitive couplings (light gray edges); mode $c$ is parametrically driven. b) Superconducting circuit implementation: modes $a$ and $b$ are transmon qubits; the coupler mode $c$ is implemented as a generalized capacitively shunted flux qubit (see text).\label{Fig:Model}}
  \end{figure}

A dominant source of infidelity in gate operation consists of cross-Kerr interactions, or the $ZZ$ terms in Pauli matrix notation. These terms are either static due to the connectivity of qubits, or dynamically generated by control drives. In the case of many two- and single-qubit gates, $ZZ$ terms produce spurious entanglement that cannot be mitigated by local single-qubit operations. There are active experimental efforts to reduce the effect of $ZZ$ interactions \cite{mundada_et_al_2019,ku2020suppression,sung2020realization,noguchi2020fast,kandala2020demonstration,zhao2020suppression}. Moreover, the presence of nonlocal $ZZ$ interactions, and of higher-order cross-Kerr terms, can indicate the onset of quantum chaotic behavior in systems of many coupled qubits \cite{berke2020transmon}.

In this paper, we present a computationally efficient set of analytical and numerical tools to characterize and tailor gate Hamiltonians. As an example application of these tools, we consider flux-tunable parametric coupler architectures \cite{mckay_et_al_2016,collodo2020implementation} schematically illustrated in \cref{Fig:Model}. We develop two complementary approaches, both of which start from a treatment of the Floquet Hamiltonian which can capture non-RWA effects exactly \cite{shirley_1965,sambe_1973,grifoni_haenggi_1998}. Our analytical approach starts from the quantization of the \textit{driven} superconducting circuit. More specifically, while we adopt a normal-mode picture such as in black-box quantization \cite{Nigg_2012} or energy-participation-ratio approaches \cite{minev2020energyparticipation}, 
the mode frequencies and impedances, and as a result 
self- and cross-Kerr interactions, depend on the strength of the drive. This dependence is accounted for in an expansion over the harmonics of the drive. From this, we obtain accurate expressions of ac-Stark shifted transition frequencies and interaction strengths. Importantly, as compared to previous work, normal modes are defined here by taking into account drive-induced corrections \cite{verney_et_al_2019} to the Josephson potential \cite{petrescu_et_al_2021}. Due to its similarity to black-box quantization, this analytical technique can be easily generalized to circuits containing multiple qubits and couplers.

To obtain corrections to the effective interaction strengths, our approach relies on a 
time-dependent Schrieffer-Wolff perturbation theory \cite{theis_wilhelm_2017,petrescu_et_al_2020,malekakhlagh_et_al_2020-2}, which consists of a hierarchy of unitary transformations applied to the time-dependent Floquet Hamiltonian \cite{malekakhlagh_et_al_2020,petrescu_et_al_2020}. We make the explicit choice to work in the transmon limit of small anharmonicity \cite{koch_et_al_2007}, expressed in terms of the small dimensionless parameter  $\sqrt{8 E_\text{C} / E_\text{J}}$, whereas drive effects are included in a series expansion over the harmonics of the drive frequency and then integrated into the exact treatment of the normal-mode Hamiltonian. This approach allows us to identify the contribution of each driven normal mode to
the different effective interaction constants. 

Our formalism is equally applicable to strong anharmonicities, where one has to formulate the Hamiltonian in the energy eigenbasis. The cross-resonance gate \cite{rigetti_devoret_2010,chow_et_al_2011} has been accurately modeled \cite{malekakhlagh_et_al_2020-2} with such methods, with the notable difference that drive effects were included in the perturbative expansion, something which requires the calculation of higher-order corrections as the drive strength is increased. In contrast, here we show that by effectively performing a series resummation over drive-amplitude contributions, we can model effects such as gate-rate saturation with drive power that are frequently observed (see e.g.~Refs.~\cite{chow_et_al_2011,Sheldon_2016}) without the need to evaluate high-order terms in perturbation theory.

On the other hand, with our numerical approach, we show how gate parameters and, more precisely, the data from a two-tone spectroscopy experiment, can be extracted from a solution of the Floquet eigenproblem \cite{shirley_1965,sambe_1973}. This is  efficient by comparison to the simulation of Hamiltonian dynamics over the full duration of the gate protocol: Floquet methods rely on integrating the dynamics over one period of the parametric drive, on the order of $1~\text{ns}$, which is typically three orders of magnitude shorter than the gate duration. By construction, the parameters extracted from this approach account for renormalization by the drives. We are then able to benchmark the convergence of the analytical approach by direct comparison to the numerical result. In the context of superconducting circuit architectures, Floquet numerical methods have also been used to model instabilities in transmon qubits under strong drives \cite{verney_et_al_2019}, to obtain corrections beyond linear-response theory for the bilinear interaction between two cavities mediated by a driven ancilla \cite{zhang_et_al_2019}, to accurately model a strongly-driven controlled-phase gate between transmon qubits \cite{krinner2020demonstration}, or to enhance the coherence of fluxonium qubits \cite{mundada_et_al_2020,huang_et_al_2021}.

The remainder of this paper is structured as follows. In \cref{Sec:Model} we introduce the circuit model, as well as a pedagogical toy model from which all qualitative features of the full theory can be extracted, and illustrate how to obtain the different gate Hamiltonians.  In \cref{Sec:PT} we introduce the basic concepts for second-order RWA, based on a Schrieffer-Wolff transformation of the Floquet Hamiltonian. \Cref{Sec:CircNOJA} captures in more detail the complexity of the problem with an analysis of the three-mode theory derived from the full-circuit Hamiltonian. In \cref{Sec:Floq}, we describe in detail a method to extract effective gate Hamiltonians from a Floquet analysis. In \cref{Sec:FullNum}, we compare all previous approaches 
using simulations based on the numerical integration of the Schr\"odinger equation. Finally, we summarize in \cref{Sec:Conc}. 

\section{Model Hamiltonian}
\label{Sec:Model}

As a concrete example of our approach, we consider a model for a parametric coupler consisting of three nonlinear bosonic modes interacting capacitively~\cite{mckay_et_al_2016}, see Fig.~\ref{Fig:Model}a). The qubit modes $\ha$ and $\hb$ are assumed to be far-detuned, making the beam-splitter (or $i$SWAP) qubit-qubit interactions negligible in the rotating-wave approximation. Those modes are capacitively coupled to a third mode, the coupler $\hc$. The latter can be parametrically modulated in order to activate interactions between the two qubit modes, for example a $i$SWAP-type gate on which we mostly focus here. 

\subsection{Superconducting circuit}

A possible realization of this three-mode system is shown in Fig.~\ref{Fig:Model}b) and consists of two fixed-frequency transmon qubits interacting via a capacitively shunted flux qubit whose two branches contain one and $N$ Josephson junctions, respectively \cite{you_et_al_2007,Steffen_2010,Yan:2016aa}. In a single-mode approximation, this generalized flux qubit plays the role of coupler mode and the parametric drive is realized by modulating the reduced external flux $\phx = 2\pi \Phi_\mathrm{ext}/\Phi_0$, with $\Phi_\mathrm{ext}$ the flux threading the coupler's loop and $\Phi_0$ the flux quantum. 
For certain values of the static external flux, the coupler has a positive anharmonicity, which is important in obtaining gates with a vanishing $ZZ$ interaction \cite{ku2020suppression,zhao2020suppression,xu2020zz,zhao_et_al_2020-2}. We stress that we use this specific circuit implementation for illustration purposes only, and that the methods presented here apply beyond the weakly-anharmonic regime.

Quantizing the circuit of Fig.~\ref{Fig:Model}c) using the standard approach \cite{fluctuations_quantiques_1997,Vool_2017} yields the Hamiltonian (see \cref{App:Quantization} for a detailed derivation)
\begin{align}\label{Eq:Model}
 \hH(t) = \hH_a + \hH_b + \hH_c(t) + \hH_g,
\end{align}
where the transmons and the coupler are described by

\begin{align}\label{Eq:CircuitHabc}
  \begin{split}
  \hH_j = 4 E_{\text{C}j} \bnj^2 &- E_{\text{J}j} \cos(\bphj), \; j= a,b,  \\
  \hH_c(t) =  4 \ECc \bnc^2 &- \alpha \EJc \cos\left[\bphc + \mu_\alpha \phx(t)\right]  \\  &- \beta N  \EJc \cos\left[ \frac{\bphc}{N}  +\mu_\beta \phx(t) \right].
  \end{split}
\end{align} 
These expressions use pairs of canonically conjugate superconducting phase difference and Cooper pair number for the bare modes, $\left[ \bphj, \bnk \right]  = i\delta_{jk}$ for the mode indices $j,k = a,b,$ or $c$, and we set $\hbar=1$. The Josephson energies are denoted $\EJa$,  $\EJb$ for the transmon modes, whereas $\beta \EJc$ is the Josephson energy of one of $N$ array junctions in the coupler, and $\alpha$ is a factor parametrizing the anisotropy between the two branches. The parameter $\beta$ is a renormalization of the superinductance due to disorder in the junction array and finite zero-point fluctuations (see \cref{App:Quantization}). Moreover, the parameter $\alpha$ accounts for a renormalization of the small junction energy due to hybridization with the modes in the junction array. Furthermore, $\ECa$, $\ECb$ and $\ECc$ are charging energies. In the transmon regime, $\EJa/\ECa$ and $\EJb/\ECb \gtrsim 50$ \cite{koch_et_al_2007}.

The coupler loop is threaded by an external flux $\phx$ which can be modulated in time with a modulation amplitude $\delta \varphi$, taken to be small compared to the flux quantum
\begin{align} \label{Eq:PhiExt}
  \phx(t) = \bphx + \delta \varphi \sin(\wdr t).
\end{align}
As discussed by \textcite{you_et_al_2019}, quantization of the coupler loop under time-dependent flux imposes that the external flux be included in both branches of the potential energy in $\hH_c(t)$, with weighting factors $\mu_{\alpha,\beta}$ determined by the capacitive energies of the two branches (see \cref{App:Quantization} for a detailed derivation). This subtlety is important, as the details of the flux modulation determine the parametric interactions between the two qubit modes. 

Finally, the three bare modes interact through linear terms induced by the capacitive coupling
\begin{align}\label{eq:Hg}
  \hH_g = 4 \ECab \bna \bnb + 4 \ECbc \bnb \bnc + 4 \ECca \bnc \bna.
\end{align}
The introduction of normal modes will eliminate this linear coupling Hamiltonian.

\subsection{Toy model for circuit Hamiltonian}
\label{Sec: KNO}
In this subsection, we introduce a simple model which captures the essential qualitative features of the Hamiltonian of~\cref{Eq:Model}. Our toy model consists of three linearly coupled Kerr-nonlinear oscillators and has the form given in \cref{Eq:Model} now with
\begin{align}
  \label{Eq:HTriangle} 
  \begin{split}
  \hH_a &= \Wa \hba^\dagger \hba + \frac{\bala}{2} \hba^{\dagger 2} \hba^2,  \\
  \hH_b &= \Wb \hbb^\dagger \hbb + \frac{\balb}{2} \hbb^{\dagger 2} \hbb^2,  \\
  \hH_c(t) &= \Wc(t) \hbc^\dagger \hbc + \frac{\balc}{2} \hbc^{\dagger 2} \hbc^2.  \\
  \hH_g    &= -\bm{g}_{ab} \hba^\dagger \hbb - \bm{g}_{bc} \hbb^\dagger \hbc - \bm{g}_{ca} \hbc^\dagger \hba +\text{H.c.}  
  \end{split} 
\end{align}
Comparing to the full-circuit model, note that $\bm{\omega}_{a(b)} \approx \sqrt{8 E_{\text{J}a(b)} E_{\text{C}a(b)}} - E_{\text{C}a(b)}$ whereas the anharmonicities of the transmon qubits are negative and amount to $\bm{\alpha}_{a(b)} \approx -E_{\text{C}a(b)}$. In an experimental implementation, the parameters defining the coupler---the anharmonicity $\balc$ and the frequency $\Wc(t)$---can be varied by applying a time-dependent external flux to activate a chosen gate. 

The parametric drive resulting from the flux modulation of \cref{Eq:PhiExt} is modeled by a modulation of the coupler frequency at a frequency $\wdr$
\begin{align}
  \Wc(t) = \Wc + \delta \sin(\wdr t). \label{Eq:wctSimple}
\end{align}
In a more detailed analysis of the coupler (see \cref{Sec:CircNOJA}), we take into account the time dependence of the anharmonicity $\balc$, but we choose to neglect it in this toy model. 

Note that we have reduced the complexity of the problem in a few ways: We have truncated the Josephson expansion to include only quartic terms. All photon number nonconserving terms have been dropped. Higher harmonics of the drive of \cref{Eq:wctSimple} are neglected, and we have not considered the ac-Stark shifts of the various coupling constants. All of these refinements are taken into account in the analysis of the full circuit Hamiltonian in \cref{Sec:CircNOJA}. Thus the toy model is significantly simpler than the full circuit theory, but nonetheless still contains the necessary ingredients that allow us to illustrate the general method introduced in this paper.

\section{Perturbative expansion}
\label{Sec:PT}
In this section, we introduce a perturbative expansion to obtain successive corrections to the effective Hamiltonian in the rotating-wave approximation. To simplify the discussion, we focus on the toy model and come back to the full circuit Hamiltonian in the next section. Our approach relies on a sequence of unitary transformations amounting to a time-dependent Schrieffer-Wolff treatment of the Floquet Hamiltonian in the normal-mode representation, an approach used before to derive corrections to the lifetime of driven transmon qubits \cite{petrescu_et_al_2020,malekakhlagh_et_al_2020}. Time-dependent extensions of Schrieffer-Wolff transformations have been shown to be necessary to capture effects of drives in the dispersive regime of circuit QED  \cite{theis_wilhelm_2017}, with quantitative agreement with experiment in the analysis of the cross-resonance gate \cite{malekakhlagh_et_al_2020-2}. A notable difference from prior work on microwave-activated two-qubit gates is that, in performing a normal-mode transformation, we are able to obtain good agreement with exact numerics already at second order in perturbation theory. For example, the calculation in Ref.~\cite{malekakhlagh_et_al_2020-2} relies on an expansion in capacitive couplings and drive power, which would require us, in the setup presented here, to go to higher order (fourth) in the calculation to obtain results comparable to the normal-mode approach.

\subsection{Formalism}
As usual, our starting point is a decomposition of the system Hamiltonian into an unperturbed, exactly solvable part, and a perturbation:
\begin{align}\label{eq:Hfull}
  \hH = \hH^{(0)}(t) + \lambda \hH^{(1)}(t).
\end{align}
Here, we have introduced the dimensionless power-counting parameter $\lambda$ to keep track of the order in perturbation theory, to be set at the end of the calculation to unity, $\lambda \to 1$. Now we move to the interaction picture with respect to $\hH^{(0)}$. Letting $\hat{U}_0(t) = \mathcal{T} e^{-i \int_0^t dt' \hat{H}^{(0)}(t')}$, where $\mathcal{T}$ is the time-ordering operator, we find for the interaction-picture Floquet Hamiltonian
\begin{align}\label{Eq:HI1}
\begin{split}
  \lambda \hH_{I}^{(1)}(t) - i\partial_t &= \hat{U}_0^\dagger(t) \left[ \hH^{(0)} + \lambda\hH^{(1)}(t) - i \partial_t \right] \hat{U}_0(t)  \\
    &= \hat{U}_0^\dagger(t) \; \lambda\hH^{(1)}(t) \;\hat{U}_0(t) - i \partial_t. 
    \end{split}
\end{align}
In the above we only assume that the unperturbed time-evolution operator $\hat{U}_0(t)$ is known. \Cref{Eq:HI1} can be seen as a unitary transformation between two Floquet Hamiltonians \cite{sambe_1973}. Thus, the Floquet quasienergies corresponding to $\lambda\hH_{I}^{(1)}(t) - i\partial_t$ must be identical to those of $\hH^{(0)}+\lambda\hH^{(1)}(t) - i\partial_t$, while the eigenstates are related by 
$\hat{U}_0(t)$.

In an iterative Schrieffer-Wolff approach, we treat the operator $\lambda \hH^{(1)}{(t)}$ as a small perturbation from which we derive corrections to the known Floquet quasienergies of $\hH^{(0)}$ \cite{malekakhlagh_et_al_2020,petrescu_et_al_2020}. To this end, we consider a unitary transformation on the interaction-picture Floquet Hamiltonian $\hH_I(t) - i\partial_t \equiv \lambda \hH_I^{(1)}(t) - i\partial_t$, and the corresponding Baker-Campbell-Hausdorff (BCH) expansion in powers of the generator of this unitary, that is
\begin{align}\label{Eq:BCHAna}
  \begin{split}
    \hH_{I,\textit{eff}} - i\partial_t 
    &\equiv e^{-\hGI(t)} [\hH_I(t) - i\partial_t] e^{\hGI(t)} \\ 
    & = \hH_I - i \dot{\hat{G}}_\text{I} + [\hH_I,\hGI] -\frac{i}{2}[\dot{\hat{G}}_\text{I},\hGI]  - i\partial_t + ... 
  \end{split}
\end{align}
This equation defines the effective Hamiltonian, whose spectrum is equal (up to a desired precision in $\lambda$) to that of the original driven theory. The generator $\hGI(t)$ can be solved for iteratively in powers of $\lambda$ (see \cref{Ap:SWPT}), which allows us to perform the rotating-wave approximation order by order
\begin{align}
  \hH_{I,\textit{eff}} = \lambda \overline{\hH}_I^{(1)} + \lambda^2 \overline{\hH}_I^{(2)} + \ldots
\end{align}
where the terms on the right-hand side are defined below. 

To obtain a lowest-order term of the effective Hamiltonian, $\lambda \overline{\hH}_I^{(1)}$, we separate the interaction picture Hamiltonian into oscillatory and non-oscillatory terms with the notation 
\begin{align}
  \lambda\hH_{I}^{(1)}(t) \equiv \lambda \bHI{1}  + \lambda \tHI{1}(t), \label{Eq:Decomp1}
\end{align}
where we define the constant part of a time-dependent operator $\hat{O}(t)$ by~\cite{mirrahimi2015dynamics}
\begin{align}\label{eq:BarOpDef}
  \overline{\hat{O}} \equiv \lim_{T\to \infty} \frac{1}{T} \int_0^T dt \hat{O}(t),
\end{align}
whereas the oscillatory part of the operator is
\begin{align}\label{eq:TildeOpDef}
  \widetilde{\hat{O}}(t) \equiv \hat{O}(t) - \overline{\hat{O}}.  
\end{align}
Since the time-averaging operation removes all terms that are oscillatory in time, $\bHI{1}$ is the first-order RWA Hamiltonian~\cite{mirrahimi2015dynamics}, whereas $\tHI{1}$ is canceled by an appropriate choice of the corresponding term at order $\lambda$ in the generator. 

One can iterate this procedure at every order, collecting terms that are oscillatory and then canceling them. The second-order RWA Hamiltonian (for a derivation, see \cref{Ap:SWPT}) reads
\begin{align}
  \begin{split}
  \lambda^2\bHI{2} = & \frac{1}{i} \overline{\left[\bHI{1}, \int_0^t  \lambda \tHI{1}(t') dt'\right]}  \\ & + \frac{1}{2i}\overline{ \left[  \lambda\tHI{1}(t), \int_0^t  \lambda \tHI{1}(t') dt' \right] }. \label{Eq:bHI2}
  \end{split}
\end{align}
This form becomes analogous to the second term in the Magnus expansion \cite{goldman_dalibard_2014,mirrahimi2015dynamics} when the perturbation has a vanishing mean, \ie $\bHI{1} = 0$.

\subsection{Black-box quantization approach to the toy model}
\label{Subsec:EffGateToy}

Expressing the toy-model Hamiltonian as the sum of static quadratic terms, $\hH^{(0)}$, and of time-dependent and Kerr terms, $\hH^{(1)}(t)$, the first step in deriving parametrically activated interactions between the transmon modes is to diagonalize the former, which we write as
\begin{align}
\hH^{(0)} = \left( \begin{array}{ccc} \hba^\dagger & \hbb^\dagger & \hbc^\dagger \end{array} \right) 
  \left( \begin{array}{ccc}
    \Wa & \bm{g}_{ab}     & \bm{g}_{ca} \\
    \bm{g}_{ab}    & \Wb  & \bm{g}_{bc} \\
    \bm{g}_{ca}    &  \bm{g}_{bc}   & \Wc  
    \end{array}
  \right)
  \left( 
  \begin{array}{c}
    \hba \\
    \hbb \\
    \hbc
  \end{array}
  \right).
\end{align}
This diagonalization is achieved with an orthonormal transformation $\hat{\bm{\alpha}} = \sum_{\beta=a,b,c} u_{\alpha\beta} \hat{\beta}$, for $\alpha=a,b,c$, and which is chosen such that $\hH^{(0)}$ takes the form
\begin{align}
  \hH^{(0)} = \wa \ha^\dagger \ha + \wb \hb^\dagger \hb + \wc \hc^\dagger \hc, \label{Eq:H0Toy}
\end{align}
where $\ha$, $\hb$ and $\hc$ are the normal modes and $\omega_{a,b,c}$ the corresponding mode frequencies. The $u_{\alpha\beta}$ are hybridization coefficients encoding the connectivity of the three modes through the capacitive couplings $\hH_g$ entering in $\hH^{(0)}$. In this normal-mode basis, the remainder of the Hamiltonian reads
\begin{align}
\begin{split} \label{Eq:H1Toy}
  &\lambda \hH^{(1)}(t)  = \\ & \sum_{j=a,b,c}  \frac{\bm{\alpha}_j}{2} (u_{ja} \ha + u_{jb} \hb + u_{jc} \hc)^{\dagger 2} (u_{ja} \ha + u_{jb} \hb + u_{jc} \hc)^2  \\
  &+\delta\sin(\wdr t)(\uca \ha + \ucb \hb + \ucc \hc)^\dagger (\uca \ha + \ucb \hb + \ucc \hc ). 
  \end{split}
\end{align}
The expression above illustrates that coupling between the normal modes arises from the nonlinearity and the parametric drive.

\begin{figure}[t!]
  \includegraphics[width=\linewidth]{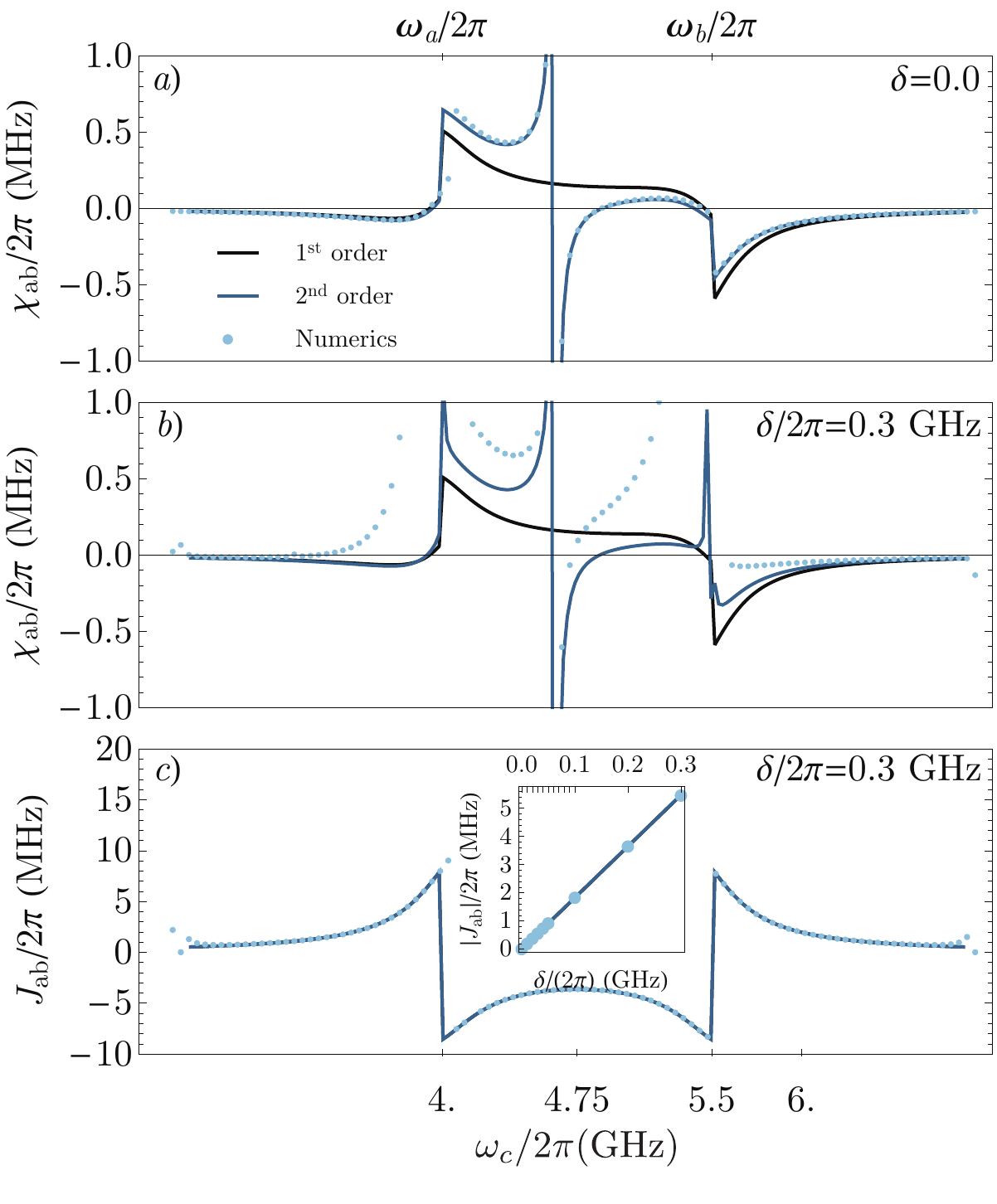}
   \caption{a) Static cross-Kerr interaction $\chi_{ab}(\Wc)$, from first (black) and second-order RWA (blue), and from the full diagonalization of \cref{Sec:Floq}) (light blue points) for: $\Wa/2\pi = 4.0$, $\Wb/2\pi = 5.5$, $\bala/2\pi=-0.3$, $\balb/2\pi=-0.2$, $\balc/2\pi =  0.25$, $\bm{g}_{ab}/2\pi=0.12 $, $\bm{g}_{bc}/2\pi = -0.12$, all in GHz, and $\bm{g}_{ca}/2\pi = 0$. b) Analogue of a) for \textit{dynamical} cross-Kerr interaction at $\delta/2\pi=0.3\, \text{GHz}$. c) Same as b) for the gate interaction rate $J_{ab}(\Wc)$. Inset: $J_{ab}(\delta)$  at $\Wc/2\pi= 4.25$~GHz.\label{Fig:JabToy}}
\end{figure}

Our choice of unperturbed Hamiltonian $\hH^{(0)}$ and perturbation $\lambda \hH^{(1)}$ in \cref{Eq:H0Toy,Eq:H1Toy}, respectively, is guided by black-box quantization \cite{Nigg_2012}: the unperturbed Hamiltonian is linear and diagonal in the normal-mode basis, whereas the perturbation consists of Kerr-nonlinear terms, on the one hand, and quadratics appearing from the normal-mode expansion of the parametric drive, on the other hand. As we show below, while better choices are possible (see \cref{Sec:AnhCorrections}), this choice leads to a simple and intuitive form for the effective Hamiltonian.

As an example of the many common types of interactions that can be activated by a parametric drive \cite{blais_et_al_2021}, an $i$SWAP interaction between the transmon modes arises if we set the modulation to be at the frequency difference between the two transmon modes
\begin{align}
  \wdr = \wb - \wa. \label{Eq:ParDriveSWAP}
\end{align}
This choice yields the first-order RWA Hamiltonian
\begin{align}\label{Eq:TMHI1}
  \begin{split}
    \lambda\bHI{1} &= J^{(1)}_{ab}\left( - i \ha^\dagger \hb + \text{H.c.}\right) \\
    &+ \frac{\ala^{(1)}}{2} \ha^{\dagger 2} \ha^2 + \frac{\alb^{(1)}}{2} \hb^{\dagger 2} \hb^2 + \frac{\alc^{(1)}}{2} \hc^{\dagger 2} \hc^2 \\
    &+ \chi_{ab}^{(1)} \ha^\dagger \ha \hb^\dagger \hb + \chi_{bc}^{(1)} \hb^\dagger \hb \hc^\dagger \hc + \chi_{ca}^{(1)} \hc^\dagger \hc \ha^\dagger \ha. 
  \end{split}
\end{align} 
The first row of this equation contains the $i$SWAP interaction of amplitude $J^{(1)}_{ab}$. The second row contains the mode anharmonicities, and the third row contains cross-Kerr interactions, the first of which is the $ZZ$ term. 

The coupling constants in the above effective Hamiltonian result from the normal-mode transformation of the quadratic part of the toy model and take the form
\begin{align}\label{Eq:TMCCsHI1}
  \begin{split}
    J_{ab}^{(1)} = & \uca \ucb \frac{\delta}{2}, \quad \alpha_j^{(1)} = \sum_{i=a,b,c} u_{ij}^4 \bm{\alpha}_i, \\
    &\chi^{(1)}_{jk} = \sum_{i=a,b,c} 2 u_{ij}^2 u_{ik}^2 \bm{\alpha}_i,       
  \end{split}
\end{align}
for all $j,k=a,b,c$, and $j \neq k$. In practice, one wants to maximize $J_{ab}^{(1)}$ to obtain a fast gate, while minimizing the cross-Kerr interactions $\chi^{(1)}_{jk}$ to avoid the accumulation of coherent errors. Cross-Kerr interactions are a source of infidelity for a $i$SWAP-type gate, as well as in other gate implementations \cite{ku2020suppression,sung2020realization,noguchi2020fast,kandala2020demonstration,zhao2020suppression,krinner_et_al_2020}. In the first-order RWA Hamiltonian, to cancel the cross-Kerr interaction between the two transmons, we use a coupler with a positive anharmonicity \cite{zhao_et_al_2020-2} $\balc > 0$, together with $\bala,\balb < 0$, which is distinct from using qubits of opposite anharmonicities \cite{ku2020suppression,xu2020zz}, or couplers with negative anharmonicity \cite{zhao2020suppression,sete2021floating}. \Cref{Eq:TMCCsHI1} forms the basis for the optimization of the gate parameters. Before pursuing this further, we first derive important corrections to the gate Hamiltonian from the oscillatory part of the Hamiltonian, $\tHI{1}(t)$. Finally, note that the first-order term $\chi^{(1)}_{jk}$ is only a \textit{static}, i.e.~$\delta$-independent, cross-Kerr interaction. 

At second order in perturbation theory, there is no correction to the $i$SWAP gate frequency $J_{ab}^{(2)} = 0$. In the regimes of interest, where the coupler frequency is close enough to the qubit frequencies for the interaction between the coupler and the qubits to be non-negligible, the dominant contribution to the second-order RWA correction to the cross-Kerr interaction $\chi_{ab}^{(2)}$ is
\begin{align}\label{Eq:Chi2RWAToy}
  \begin{split} 
    \chi_{ab}^{(2)} & \approx 2 \frac{\left(\sum_{j=a,b,c} u_{aj} u_{bj} u_{cj}^2 \balj \right)^2}{\wa + \wb - 2 \wc} \\
  &+ \delta\frac{\uac \ubc \left[\uaa^3 \uba \bala -\ubb^3 \uab \balb \right]}{\wa-\wb}.
  \end{split}
\end{align}
The full expression for $\chi_{ab}^{(2)}$ can be found in \cref{Ap:PTToy}.
Inspecting the hybridization coefficients $u_{\alpha\beta}$ and the denominators in \cref{Eq:Chi2RWAToy}, we deduce that the second-order correction to the \textit{static} cross-Kerr interaction, corresponding to the first term in \cref{Eq:Chi2RWAToy}, arises from a virtual two-photon excitation of the coupler (generated by the commutator $[\ha \hb \hc^{\dagger 2}, \ha^\dagger \hb^\dagger \hc^2]$). This correction would not be present in a two-level approximation \cite{mckay_et_al_2016}. On the other hand, the second term in \cref{Eq:Chi2RWAToy} is the lowest-order contribution to the \textit{dynamical} cross-Kerr interaction.

\subsection{Improving the starting point of the perturbation theory}
\label{Sec:AnhCorrections}

As mentioned in the previous subsection, other choices for $\hH^{(0)}$ and $\lambda\hH^{(1)}$ are possible which give better accuracy in comparisons with exact Floquet numerics. In this subsection, we take the unperturbed Hamiltonian $\hH^{(0)}(t)$ to consist of the Fock-space diagonal part of $\hH(t)$, namely
\begin{align}\label{Eq:H0RefToy}
  \begin{split}
    \hH^{(0)}(t) &= \wa \ha^\dagger \ha + \wb \hb^\dagger \hb + \wc \hc^\dagger \hc  \\
    &  + \delta \sin(\wdr t) \left[ \uca^2 \ha^\dagger \ha + \ucb^2 \hb^\dagger \hb + \ucc^2 \hc^\dagger \hc \right] \\
    &  +\frac{\ala^{(0)}}{2} \ha^{\dagger 2}\ha^2+\frac{\alb^{(0)}}{2}  \hb^{\dagger 2}\hb^2 + \frac{\alc^{(0)}}{2}  \hc^{\dagger 2}\hc^2  \\
    & + \chi_{ab}^{(0)} \ha^\dagger \ha \hb^\dagger \hb + \chi_{ac}^{(0)} \ha^\dagger \ha \hc^\dagger \hc + \chi_{bc}^{(0)} \hb^\dagger \hb \hc^\dagger \hc,   
  \end{split}
\end{align}
where the quartic couplings in the last two rows are exactly those defined in \cref{Eq:TMCCsHI1}, with the superscript now changed from $1$ to $0$ to reflect their presence in the unperturbed Hamiltonian.

We expect this starting point, \cref{Eq:H0RefToy}, to lead to more precise results, because of two reasons. Firstly, the perturbation $\hH^{(1)}$ is now off-diagonal in Fock-space. The effects of the anharmonicities of the modes are now included at the level of $\hH^{(0)}$, and, in particular, we expect a dressing of contributions corresponding to two-photon excitations, such as \cref{Eq:Chi2RWAToy}. Secondly, due to the second row of \cref{Eq:H0RefToy}, we can derive the effect of harmonics of the drive frequency through a Fourier expansion of the time-evolution operator $\hat{U}_0(t)$, defined below.

Following the steps of the previous subsections, we 
evaluate the time evolution operator with respect to the unperturbed Hamiltonian $\hH^{(0)}(t)$, that is
\begin{align}
  \hat{U}_0(t) = e^{-i \int_0^t dt' \hH^{(0)}(t')},
\end{align} 
with $[\hH^{(0)}(t),\hH^{(0)}(t')]=0$.  The time-dependence in the exponent is handled via the Jacobi-Anger identity~\cite{Abramowitz_Handbook_1964} 
\begin{align}
  e^{i \hat{O} \frac{\delta}{\omega_d} \cos\wdr t} =J_0\left(\frac{\hat{O}\delta}{\wdr} \right) +  2\sum_{n=1}^{\infty} i^n J_n\left(\frac{\hat{O}\delta}{\wdr}\right) \cos n \wdr t,
\end{align}
for any operator $\hat{O}$, where $J_n(z)$ is the $n^{\textit{th}}$ Bessel function of the first kind~\cite{Abramowitz_Handbook_1964}.  This expansion allows us to keep track of all harmonics of the drive. In practice, since the modulation amplitude is small, $\delta/\wdr \ll 1$, only a few terms will be necessary.

We obtain to first order (and truncating after the first Bessel function)
\begin{align}
\begin{split}
  J_{ab}^{(1)} = \frac{\delta}{2} \uca \ucb \Bigg[ &J_0\left(\frac{\delta \uca^2}{\wdr}\right) J_0 \left(\frac{\delta \ucb^2}{\wdr}\right) \\ 
   &+ 3  J_1\left(\frac{\delta \uca^2}{\wdr}\right) J_1\left(\frac{\delta \ucb^2}{\wdr}\right)\Bigg],
\end{split}
\end{align}
which agrees, up to linear terms in $\delta$, with the expression in \cref{Eq:TMCCsHI1}. On the other hand, the cross-Kerr interaction at this order is vanishing $\chi_{ab}^{(1)} = 0$.

To second order in perturbation theory, the dominant contributions to $J_{ab}^{(2)}$ are
\begin{align}
  \begin{split}
    J_{ab}^{(2)} =  \frac{i \delta^2 }{2}& \uca \ucb \ucc^2\left(\frac{1}{\wa-\wc} - \frac{1}{\wb - \wc} \right) \\ 
    & \times J_1\left(\frac{\delta \ucc^2}{\wdr}\right)  \prod_{j=a,b,c} J_0\left( \frac{\delta u_{cj}^2}{\wdr}\right) + \ldots.
  \end{split}
\end{align} 
We do not reproduce here the full form containing 20 terms. The second-order contribution $\chi_{ab}^{(2)}$ is non-vanishing, and contains approximately 450 terms in expanded form.  
Despite the complexity of these full expressions, they are easy to derive and manipulate with symbolic computation tools \cite{ZITKO20112259}.
Focusing on the static cross-Kerr interaction, \ie in the $\delta \to 0$ limit, the dominant correction resulting from the above changes amounts to our previous \cref{Eq:Chi2RWAToy}, but replacing the denominator of the first term of that expression by a form that faithfully includes the contribution from the anharmonicities, as expected in the case of a virtual two-photon excitation of the coupler mode. That is, approximately
\begin{align}
  \wa+ \wb - 2 \wc \to \wa+ \wb - (2 \wc + \ucc^4 \balc).
\end{align}  
The full expression for the corrected static cross-Kerr interaction can be found in \cref{Ap:PTToy}. 

In \cref{Fig:JabToy} we compare analytical results to numerical results obtained from exact diagonalization (at $\delta=0$), or a solution of the Floquet eigenspectrum at $\delta \neq 0$ (see \cref{Sec:Floq}). We find that the agreement between numerics and analytics is excellent for the gate interaction strength, as well as for the static $\delta=0$ cross-Kerr interaction. However, we find that second-order perturbation theory is insufficient to reproduce the effects of the drive on the cross-Kerr interaction, even for modest drive amplitudes. We expect that higher-order perturbation theory should correctly capture the drive-amplitude dependence of the anharmonicities, but these contributions have been inaccessible in our study due to the large memory demands of the computer algebra manipulations.

\section{Full circuit Hamiltonian}
\label{Sec:CircNOJA}

Building on the previous results, we now turn to deriving an effective Hamiltonian for the full circuit Hamiltonian of \cref{Eq:Model} and Eq.~(\ref{Eq:CircuitHabc}). 
The full circuit model goes beyond the toy model in that it systematically includes the effects of the parametric drive on all of the coupling constants. Although the simplicity of the toy model is useful in developing an intuitive understanding of the effect of parametric drives on the system, the full circuit model can lead to more accurate comparisons with experimental data.

The full circuit theory is constructed with the following steps: We first introduce creation and annihilation operators for the bare circuit modes starting from the first-order RWA driven circuit Hamiltonian in \cref{Sec:BareModeNumSim}. Because the drive is taken into account at that level, the frequencies and zero-point fluctuations of these bare modes will be explicitly corrected by the drive. In \cref{Sec: eff_gate_H}, we perform a normal-mode transformation amounting to a \textit{driven} black-box quantization approach. We then show in \cref{Sec:OtherGates} how a variety of quantum gates can be addressed by appropriate choices of the parametric drive frequency. Lastly, we find corrections to the desired gate Hamiltonian using a time-dependent Schrieffer-Wolff perturbation theory in \cref{Sec:FullCircPT}.

\subsection{Bare-mode Hamiltonian}
\label{Sec:BareModeNumSim}

To define the bare modes, we begin with the full circuit model Hamiltonian of~\cref{Eq:Model,Eq:CircuitHabc}. We normal-order expand the Josephson cosine potentials in this Hamiltonian over a set of creation and annihilation operators, which we define as follows
\begin{align}\label{Eq:coordinates}
  \begin{split}
    \bpha &= \sqrt{\frac{\eta_a}{2}} (\hba + \hba^\dagger),\\
    \bna &= -i \sqrt{\frac{1}{2\eta_a}}(\hba - \hba^\dagger),  
  \end{split}
\end{align}
with analogous equations for modes $\hbb$ and $\hbc$. The coefficients $\eta_{a,b,c}$ are chosen such that terms proportional to $\hba^2,\hbb^2$, and $\hbc^2$ vanish in the time-averaged Hamiltonian. This amounts to three transcendental equations:
\begin{align}\label{eq:transcendentalEta}
  \mathcal{F}(\eta_{j}) \eta_{j}^2  = 8 E_{\text{C}j}/E_{\text{J}j},
\end{align}
for $j=a,b, c$, where we have defined the form factors
\begin{align}
  \begin{split}
  \mathcal{F}(\eta_{a(b)}) &\equiv e^{-\frac{\eta_{a(b)}}{4}}, \\
  \mathcal{F}(\eta_{c})   &\equiv  \alpha   e^{-\frac{\eta_c}{4}}  J_0(\delta ) \cos \left(\bphx\right)  + \beta e^{-\frac{\eta_c}{4 N^2}}/N. 
  \end{split}
\end{align}
Two remarks are in order: First, in the transmon limit $\mathcal{F}(\eta_{a(b)}) \approx 1$ where we recover the usual expression $\eta_{a(b)} \approx \sqrt{8 E_{\text{C}a/b}/E_{\text{J}a/b}}$. Second, the parameter $\eta_c$ depends on the parametric drive amplitude $\delta$, which indicates that the mode $c$ impedance is drive-dependent. This has important consequences for the precision of the calculation of coupling constants dressed by the parametric drives. In particular it allows us to capture the ac-Stark shift of the coupler mode at the lowest order in perturbation theory. In what follows, sine and cosine functions of the phase are normal-order expanded according to \cref{Eq:CosSinNO} of \cref{Ap:FullCircuitTheory}. In turn, trigonometric functions of the flux modulation are expanded in Jacobi-Anger series over the harmonics of the frequency of the drive. 

Using the above definitions, the transmon Hamiltonian $\hH_a$ takes the familiar form
\begin{align}
  \hH_a = \Wa \hba^\dagger \hba - \EJa \left(\cos  \bpha  + e^{-\frac{\eta_a}{4}} \frac{\bpha^2}{2}\right). \label{Eq:HaSpider}
\end{align}
The second term on the right-hand side contains the nonlinear part of the Josephson potential, \ie the inductive part is subtracted. Up to quartic order, $\hH_a$ takes the form
\begin{align}\label{Eq:HaQuartic}
  \begin{split}
    \hH_a &= \Wa \hba^\dagger \hba + \frac{\bala}{2} \hba^{\dagger 2} \hba^2  \\
    & + \frac{\bala}{12} \left( \hba^4+\hba^{\dagger 4}\right) + \frac{\bala}{3} \left( \hba^\dagger \hba^3+ \hba^{\dagger 3} \hba \right) + \cdots
  \end{split}
\end{align}
The first row of this expression is a Kerr oscillator Hamiltonian as in the toy model of \cref{Sec:Model}, whereas the second row contains corrections from quartic counter-rotating terms. Here, we have introduced the mode frequency and anharmonicity which take the forms~\cite{koch_et_al_2007}
\begin{align}
  \begin{split}
    \Wa &= \frac{4 \ECa}{\eta_{\mathit{a}}}+\frac{1}{2} \mathcal{F}(\eta_a) \eta_a
   \EJa \approx \sqrt{8 \ECa \EJa} - \ECa,\\ 
   \bala &= -\ECa.
  \end{split}
\end{align}
Note that for the approximate equality in the first row we have used a Taylor expansion of \cref{eq:transcendentalEta} for $\eta_a$. The equations for mode $\hbb$ are identical from the above with a change of subscripts and operators $a \to b$.

The coupler Hamiltonian differs from that of the transmon modes in two fundamental ways: It breaks parity symmetry due to the external flux, and it is time-dependent. Following \cref{eq:TildeOpDef,eq:BarOpDef}, we write this time-dependent Hamiltonian as
\begin{align}\label{eq:HcBarPlusTilde}
  \hH_c(t) = \overline{\hH}_c(t) + \widetilde{\hH}_c(t).
\end{align}
The creation and annihilation operators of the coupler mode can then be defined by extracting the quadratic part of the time-averaged coupler Hamiltonian. Using \cref{Eq:coordinates} where $a\rightarrow c$ together with
\begin{align}
  \overline{\cos\left[\bphc + \mu_\alpha \phx(t)\right]} = \cos(\mu_\alpha \bphx) J_0(\mu_\alpha \delta) \cos(\bphc)
\end{align}
and a similar relation for the second branch of the coupler [see \cref{Eq:CircuitHabc}], we find in analogy to \cref{Eq:HaSpider} for the Hamiltonian of the transmon mode
\begin{align}\label{Eq:barHc}
  \begin{split}
  &\overline{\hH}_c =  \Wc \hbc^\dagger \hbc\\  
  &-\alpha \EJc J_0(\mu_\alpha\delta) \cos(\mu_\alpha\bphx) \left(\cos\bphc + e^{-\frac{\eta_c}{4}} \frac{\bphc^2}{2} \right)  \\
  &- \beta N  \EJc J_0(\mu_\beta\delta) \cos(\mu_\beta\bphx)   \left(\cos \frac{\bphc}{N} + e^{-\frac{\eta_c}{4 N^2}}\frac{\bphc^2}{2N^2}  \right)  \\
  &+ \alpha \EJc J_0(\mu_\alpha\delta)
   \sin(\mu_\alpha\bphx) \sin\bphc\\
  &+ \beta N  \EJc J_0(\mu_\beta\delta) \sin(\mu_\beta\bphx)   \sin \frac{\bphc}{N}.
  \end{split}
\end{align}
Crucially, in this first-order rotating-wave approximation of the parametric drive, the Josephson energy is renormalized by the factor $J_0(\mu_{\alpha,\beta}\delta)$, see also \cite{verney_et_al_2019}. We interpret this as an effective reduction of the Josephson potential barrier, and consequently an increase of phase fluctuations, in the presence of drives. Moreover, the presence of the non-zero external flux results in the parity breaking sine terms in \cref{Eq:barHc}.

The second term of $\hH_c(t)$ in \cref{eq:HcBarPlusTilde}, the oscillatory part, take the form
\begin{align}
  \begin{split}
    \widetilde{\hH}_c(t) =& - \alpha \EJc  \widetilde{\cos\left[\bphc + \mu_\alpha \phx(t)\right]}  \\
  & - \beta N  \EJc  \widetilde{\cos\left[\frac{\bphc}{N} + \mu_\beta \phx(t)\right]},
  \end{split}
\end{align}
which can be expanded in a Jacobi-Anger series in harmonics oscillating at the frequency $n \wdr$, where $n$ is an integer.

As above, the next step is to expand the coupler Hamiltonian up to quartic terms in the creation and annihilation operators. In contrast to the transmon Hamiltonian of \cref{Eq:HaQuartic}, parity breaking leads to the appearance of monomials of odd order. The non-oscillatory part is
\begin{align}\label{Eq:HcQuartic}
  \begin{split}
    \overline{\hH}_c &= \Wc \hbc^\dagger \hbc + \frac{\balc}{2} \hbc^{\dagger 2} \hbc^2 \\
        &  + \frac{\balc}{12} \left( \hbc^4+\hbc^{\dagger 4}\right) + \frac{\balc}{3} \left( \hbc^\dagger \hbc^3+ \hbc^{\dagger 3} \hbc \right)  \\
        &  +  \bm{g}_{c,3} \left(  \hbc^3 + \hbc^{\dagger 3} + 3 \hbc^\dagger \hbc^2 + 3 \hbc^{\dagger 2} \hbc  \right)  \\
        &  + \bm{g}_{c,1} \left( \hbc + \hbc^\dagger \right) + \cdots. 
  \end{split}
\end{align}
The first row of the above expression takes the form of the coupler Hamiltonian in the approximation of the toy model of \cref{Sec:Model}, while the remaining rows contain counter-rotating terms to quartic order. Here, the parametric drive-dependent mode frequency and anharmonicity read
\begin{align}
  \begin{split}
    \Wc &= \frac{4 \ECc}{\eta_c} + \frac{1}{2} \mathcal{F}(\eta_c) \eta_c \EJc, \\
    \balc &= -\ECc, \\
  \end{split}
\end{align}
while the prefactors of the counter-rotating terms are
\begin{align}
  \begin{split}
    \bm{g}_{c,3} &= -\alpha  \epsilon  e^{-\frac{\eta_c}{4}} \eta_c^{3/2} J_0(\mu_\alpha \delta ) \sin \left(\mu_\alpha \bphx\right) \EJc /(12 \sqrt{2}) \\
    & -\frac{\beta}{N^2} e^{-\frac{\eta_c}{4N^2}} \eta_c^{3/2} J_0(\mu_\beta\delta ) \sin \left(\mu_\beta \bphx\right) \EJc /(12 \sqrt{2}), \\
    \bm{g}_{c,1} &= \alpha  \epsilon  e^{-\frac{\eta_c}{4}} \eta_c^{1/2} J_0(\mu_\alpha \delta ) \sin \left(\mu_\alpha \bphx\right) \EJc /\sqrt{2}  \\
    & +\beta e^{-\frac{\eta_c}{4N^2}} \eta_c^{1/2} J_0(\mu_\beta\delta ) \sin \left(\mu_\beta \bphx\right) \EJc /\sqrt{2}.
  \end{split}
\end{align}
The contribution from the oscillatory part $\widetilde{\hH}_c(t)$ is too lengthy to be reproduced here, and is given up to the second harmonic of the parametric modulation frequency $\wdr$ in \cref{Tab:TildeHcQuartic} of \cref{Ap:FullCircuitTheory}.

Finally, the last term of the full circuit Hamiltonian to consider is the linear interaction $\hH_g$ induced by the capacitive coupling. Using \cref{Eq:coordinates}, this Hamiltonian takes the form
\begin{align}
  \hH_g = -\frac{2\ECab }{\sqrt{\eta_a \eta_b}}  (\hba-\hba^\dagger) (\hbb-\hbb^\dagger) + \ldots
\label{Eq:HTriangleCircuit}
\end{align}
where the ellipsis represents two more terms corresponding to the cyclic permutations of the mode indices. 

The Hamiltonian specified by Eqs.~(\ref{Eq:HaQuartic}) and its equivalent for the $b$ transmon mode, \cref{Eq:HcQuartic}, the terms summarized in \cref{Tab:TildeHcQuartic} of \cref{Ap:FullCircuitTheory}, and \cref{Eq:HTriangleCircuit}, form the basis of both the normal-mode analysis, and of the full-circuit numerical simulation performed in \cref{Sec:Floq}.

\subsection{Driven black-box quantization approach for parametrically activated interactions}
\label{Sec: eff_gate_H}
Building upon the above results, we now follow the procedure developed with the toy model in \cref{Subsec:EffGateToy} to obtain effective gate Hamiltonians under parametric modulations. To do so, we first collect under $\hH^{(0)}$ the time-independent quadratic terms derived above. We then eliminate the linear coupling $\hH_g$ of \cref{Eq:HTriangleCircuit} from $\hH^{(0)}$ through a normal-mode transformation
\begin{align}
\begin{split}
  \hH^{(0)} &= \Wa \hba^\dagger \hba + \Wb \hbb^\dagger \hbb + \Wc \hbc^\dagger \hbc + \hH_g  \\
            &\equiv \wa \ha^\dagger \ha + \wb \hb^\dagger \hb + \wc \hc^\dagger \hc.
\end{split}
\end{align}
The linear transformation is determined by a set of 18 hybridization coefficients that relate bare mode coordinates to normal mode coordinates (see \cref{App:NormalMode} for the procedure to compute these coefficients)
\begin{align}\label{Eq:NMTransf}
\begin{split}
    \hat{\bm{\varphi}}_\alpha &= \sum_{\beta=a,b,c} \frac{u_{\alpha\beta}}{\sqrt{2}} (\hat{\beta}+\hat{\beta}^\dagger),  \\
    \hat{\bm{n}}_\alpha &= \sum_{\beta=a,b,c} \frac{v_{\alpha\beta}}{i\sqrt{2}} (\hat{\beta}-\hat{\beta}^\dagger), 
    \end{split}
\end{align}
for $\alpha = a,b,c$. We stress that the hybridization coefficients $u_{\alpha\beta}$, $v_{\alpha\beta}$ depend on the amplitude of the parametric drive. As a result, drive effects such as the ac-Stark shift of the nonlinear oscillators are accounted for already at the level of the normal-mode decomposition of the circuit Hamiltonian.

In analogy with our treatment in \cref{Sec:Model} of the toy model, we have taken $\hH^{(0)}$ to be the unperturbed Hamiltonian with respect to which  
the interaction picture is defined. Primarily in order to keep the expressions more concise, we opt to neglect the corrections analyzed in \cref{Sec:AnhCorrections}.  With $\hH^{(0)}$ the unperturbed Hamiltonian, the remaining interaction terms are $\lambda\hH^{(1)}(t) \equiv \hH - \hH^{(0)}$. Expressing these in the interaction picture with respect to $\hat{U}_0 = e^{- i \hH^{(0)} t}$ as in \cref{Eq:HI1}, we find 
\begin{align}
  \lambda\hH_I^{(1)}(t) = \hat{U}_0^\dagger(t)\left[ \hH(t) - \hH^{(0)} \right] \hat{U}_0(t), 
\end{align}
which, as before, is decomposed into oscillatory and non-oscillatory parts. 

As a first example, to realize a beam-splitter interaction, the first-order RWA Hamiltonian is obtained in the form of \cref{Eq:TMHI1} for a modulation frequency that satisfies
\begin{align}
  \wdr = \wb - \wa. \label{Eq:WdrBeamSplitter}
\end{align}
Importantly, as already mentioned, the right-hand side of the above definition depends implicitly on the drive frequency $\wdr$, since it is defined in terms of ac-Stark shifted normal-mode frequencies. In \cref{Sec:Floq} we present a numerical procedure to obtain the parametric drive frequency.  

With this choice of modulation frequency, the effective Hamiltonian takes the form
\begin{align}\label{Eq:CircuitHI1}
  \begin{split}
    \lambda\bHI{1} &= J^{(1)}_{ab} ( - i \ha^\dagger \hb + \text{H.c.})  \\
    &+ \frac{\ala^{(1)}}{2} \ha^{\dagger 2} \ha^2 + \frac{\alb^{(1)}}{2} \hb^{\dagger 2} \hb^2 + \frac{\alc^{(1)}}{2} \hc^{\dagger 2} \hc^2  \\
    &+ \chi_{ab}^{(1)} \ha^\dagger \ha \hb^\dagger \hb + \chi_{bc}^{(1)} \hb^\dagger \hb \hc^\dagger \hc + \chi_{ca}^{(1)} \hc^\dagger \hc \ha^\dagger \ha  \\
    &+ J^{(1)}_{ab;a} ( - i \ha^\dagger \ha \ha^\dagger \hb + \text{H.c.} )  \\
    &+ J^{(1)}_{ab;b} ( - i \hb^\dagger \hb \ha^\dagger \hb + \text{H.c.} )  \\
    &+ J^{(1)}_{ab;c} ( - i \hc^\dagger \hc \ha^\dagger \hb + \text{H.c.} )  \\
    &+ K^{(1)}_{ab}   ( \ha^{\dagger2} \hb^2 + \text{H.c.} ). 
  \end{split}
\end{align}
In contrast to the effective gate Hamiltonian \cref{Eq:TMHI1} obtained for the toy model, there are additional terms in the last four rows, namely photon-number-conditioned beam-splitter terms and a photon-pair beam-splitter term. The prefactors of the above Hamiltonian are
\begin{align}\label{Eq:CircuitCCsHI1}
  \begin{split}
    J_{ab}^{(1)}&=-\frac{\uca \ucb}{2} \alpha  \epsilon  J_1(\mu_\alpha \delta )  \sin(\mu_\alpha \bphx) \EJc^{(\alpha)}  \\
    & -\frac{\uca \ucb}{2} \frac{\beta}{N}  J_1(\mu_\beta \delta )  \sin(\mu_\beta \bphx) \EJc^{(\beta)}, \\
    \alpha_j^{(1)} &= -\frac{1}{8} \sum_{i=a,b,c} u_{ij}^4 E'_{\text{J},i}, \\ 
    \chi_{jk}^{(1)} &= -\frac{1}{4} \sum_{i=a,b,c} u_{ij}^2 u_{ik}^2 E_{\text{J},i}',   \\
   J_{ab;j}^{(1)}&= - \frac{u_{cj}^2}{4} J_{ab}^{(1)}, \text{ for } j=a,b,c, \\
   K^{(1)}_{ab} &= -\frac{\uca^2 \ucb^2}{16} \alpha  \epsilon  J_2(\mu_\alpha \delta)  \cos(\mu_\alpha\bphx) 
   \EJc^{(\alpha)}  \\
   & -\frac{\uca^2 \ucb^2}{16} \frac{\beta}{N^3}  J_2(\mu_\beta \delta) \cos(\mu_\beta \bphx) 
   \EJc^{(\beta)},
  \end{split}
\end{align}
where we have used
\begin{align}
  \begin{split}
    \EJa' &\equiv e^{ -\frac{\uaa^2}{4}-\frac{\uab^2}{4}-\frac{\uac^2}{4} } \EJa,  \\
    \EJb' &\equiv e^{ -\frac{\uba^2}{4}-\frac{\ubb^2}{4}-\frac{\ubc^2}{4} } \EJb,  \\
    \EJc' &\equiv \alpha  \epsilon  J_0(\mu_\alpha \delta )  \cos (\mu_\alpha \bphx) \EJc^{(\alpha)} \\
    & + (\beta/N^3) J_0(\mu_\beta \delta )  \cos (\mu_\beta \bphx) \EJc^{(\beta)},  \\
    \EJc^{(\alpha)} &\equiv e^{ -\frac{\uca^2}{4}-\frac{\ucb^2}{4}-\frac{\ucc^2}{4} } \EJc, \\
    \EJc^{(\beta)} &\equiv e^{ -\frac{\uca^2}{4N^2}-\frac{\ucb^2}{4N^2}-\frac{\ucc^2}{4N^2} } \EJc.
  \end{split}
\end{align}

The above expressions depend on the drive both explicitly, through the Bessel functions, and implicitly, through the hybridization coefficients $u_{jk}$. The first three lines of \cref{Eq:CircuitCCsHI1} are similar in form to those obtained for the toy model in Eqs.~(\ref{Eq:TMCCsHI1}). Of the two additional classes of terms possible in the full circuit model at this order in perturbation theory, the photon-pair beam-splitter term, in the last row of \cref{Eq:CircuitHI1}, is generated by the second harmonic of the drive. However, since this term is fourth order in the hybridization coefficients, it can only become comparable to the beam-splitter interaction at vanishing external flux $\bphx \approx 0$, or if $\delta$ is set to cancel $J_{ab}^{(1)}$.

As in the case of the toy model, going to second order in perturbation theory using \cref{Eq:bHI2} we find corrections to the coupling constants derived above to first order. To increase the accuracy of these calculations, we first displace the coupler phase variable by a canonical transformation $\bphc \to \bphc + \bm{\varphi}_{c,\mathrm{cls}}$, where the time-independent real number $\bm{\varphi}_{c,\mathrm{cls}}$ denotes the minimum of the static coupler potential in \cref{Eq:CircuitHabc} resulting from
\begin{align}
  \begin{split}
    &\alpha  J_0(\mu_\alpha \delta\phi) \sin\left[\bm{\varphi}_{c,\mathrm{cls}} + \mu_\alpha \bphx \right]  \\&+ \beta  J_0(\mu_\beta \delta\phi) \sin\left[ \frac{\bm{\varphi}_{c,\mathrm{cls}}}{N}  +\mu_\beta \bphx \right] = 0.
  \end{split}
\end{align}
This displacement allows us to remove all static terms which are linear in the coupler quadratures $\{\bphc, \bnc\}$, before performing the normal-mode transformation. As such, these terms are accounted for exactly, and not as part of the perturbative expansion. Secondly, noting that parity-breaking terms significantly dress the coupler $0 \to 1$ transition frequency, we absorb this renormalization into a reparametrization of the external flux $\bphx \to \bphx'(\bphx)$ such that $\wc(\bphx') = \wc^{(2)}(\bphx)$, \ie~we absorb the corrections to the coupler pole at second order in perturbation theory into a redefinition of the coupler normal mode, in a self-consistent approach that can be further validated with exact numerics. 

\begin{figure}[t!]
  \includegraphics[width=\linewidth]{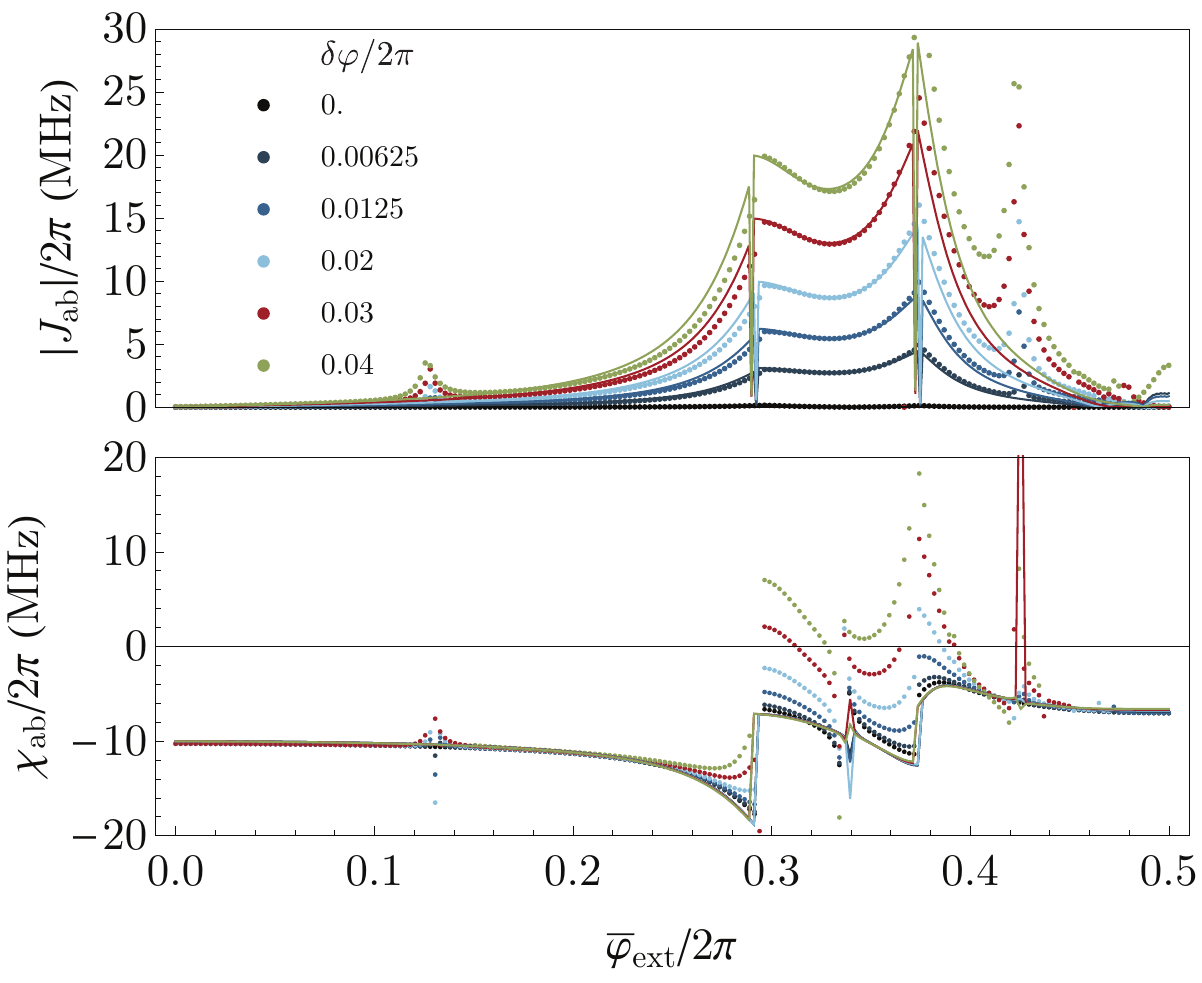}
  \caption{Coupling constants in the effective Hamiltonian for the full circuit as a function of external DC flux $\bphx$. Dots (lines) represent Floquet two-tone spectroscopy data with Hilbert space dimension 10 per mode (second-order RWA calculations). Color (see legend) encodes parametric drive amplitude $\delta \varphi / 2\pi$. Parameter choices: $C_a=134.205$ fF, $C_b=134.218$ fF, $C_c=75.987$ fF, $C_{ac}=11.11$ fF, $C_{bc}=11.22$ fF, $C_{ab}=0$, $\EJa/2\pi=37$ GHz, $\EJb/2\pi=27$ GHz, $\EJc/2\pi=50$ GHz, $\alpha=0.258,\;\beta=1$, and $N=3$. We attribute large discontinuities in the numerical curves to state tracking errors near avoided crossings (see \cref{Sec:Floq}). \label{Fig:FC_ChiJab_An_Num}}
\end{figure}

In \cref{Fig:FC_ChiJab_An_Num} we show a comparison between exact Floquet numerics (see \cref{Sec:Floq}) and second-order perturbation theory for the full circuit model. We find that the analytics reproduce with good accuracy the numerical results for the gate interaction rate $J_{ab}$ in the region where the coupler $0\to 1$ frequency lies between the two transmons: $\wa < \wc < \wa $. There are poles in the numerical gate rate $J_{ab}$ for $\wc < \wa$ or for $\wb < \wc$ that we expect to capture only at third order in perturbation theory. The numerical cross-Kerr interaction, as in the case of the toy model, only agrees well with analytics in the static case $\delta \varphi = 0$. Focusing our attention on the curves obtained from Floquet numerics, we see that with a typical set of parameters gate rates as large as $J_{ab}/2\pi \sim 20$~MHz (equivalent to a 25 ns $\sqrt{i\mathrm{SWAP}}$ gate) can be achieved while maintaining a vanishing dynamical cross-Kerr interaction. The tools presented in this paper feed into a larger scale optimization of the circuit parameters, which forms the subject of a future study.

\subsection{Other parametric gates}
\label{Sec:OtherGates}

\begin{table*}
  \begin{tabular}{|c|c|c|c|c|}
  \hline
  Gate & Bosonic operator & Drive frequency & Dominant unwanted interaction & Equation \\
  \hline 
  $i$SWAP/beam-splitter & $-i\ha^\dagger \hb + i\hb^\dagger \ha$ & $\wa - \wb$ & $\ha^\dagger \ha \hb^\dagger \hb$ & \cref{Eq:CircuitHI1} \\
  \hline
  Two-mode squeezing & $-i\ha^\dagger \hb^\dagger + i\hb \ha$ & $\wa + \wb$ & $\ha^\dagger \ha \hb^\dagger \hb$ & Eqs.~(\ref{Eq:CircuitHI1}) and (\ref{Eq:BStoTMS}) \\
  \hline
  CZ/Ising-ZZ & $\ha^\dagger \ha \hb^\dagger \hb$ & no drive &   & \cref{Eq:CircuitHI1} \\
  \hline
  CNOT & $-i(\ha-\ha^\dagger) \hb^\dagger \hb$ & $\wa$ &  $-i(\ha-\ha^\dagger) \ha^\dagger \ha$ & \cref{Eq:CrossResonance} \\
  \hline
  CSWAP & $-i\hc^\dagger \hc( \ha^\dagger \hb - \hb^\dagger \ha )$ & $\wa - \wb$ & $-i\ha^\dagger \hb + i\hb^\dagger \ha$ & \cref{Eq:CircuitHI1} \\ 
  \hline
  \end{tabular}
  \caption{\label{Tab:Gates}List of the most accessible gate Hamiltonians realizable with a parametric drive in the analyzed architecture.}
\end{table*}

The space of parametric gates is not limited to beam splitter-type, or red sideband, terms. Indeed, different interactions can be activated by appropriate choices of the frequency of the parametric drive ~\cite{bertet_et_al_2006,niskanen_et_al_2006,niskanen2007quantum,beaudoin_et_al_2012,mckay_et_al_2016,Reagor_2018,caldwell_et_al_2018,didier_et_al_2018}. For example, if instead the modulation frequency targets the blue sideband,
\begin{align}
  \wdr = \wa + \wb,
\end{align}
then the resulting interaction is a two-mode squeezing term. The effective gate Hamiltonian is formally the same as \cref{Eq:CircuitHI1} with the simple modification
\begin{align}
  \ha^\dagger \hb \to \ha^\dagger \hb^\dagger, \label{Eq:BStoTMS}
\end{align} 
in the first line and in the last four lines of \cref{Eq:CircuitHI1}. The coupling constants remain as in \cref{Eq:CircuitCCsHI1}. 

It is also possible to obtain a CNOT interaction induced by a parametric drive at $\wdr = \wa$, which makes the $a$ transmon mode into the target mode of a cross-resonance protocol \cite{rigetti_devoret_2010,chow_et_al_2011}. Following the same procedure as in the preceding subsection, with this choice of modulation frequency we arrive at the effective gate Hamiltonian
\begin{align}\label{Eq:CrossResonance}
  \begin{split}
    \lambda \bHI{1} &= -i\Omega_{a;b} (\ha - \ha^\dagger) \hb^\dagger \hb  -i\Omega_{a;c} (\ha - \ha^\dagger) \hc^\dagger \hc\\
    & -i\Omega_a (\ha - \ha^\dagger) -i\Omega_{a;a} (\ha^\dagger \ha \ha - \ha^\dagger \ha^\dagger \ha)
  \end{split}
\end{align}
The first term of above expression generates the cross-resonance gate, while the second term is a coupler-state conditional drive on mode $a$ which is negligible for
$\langle \hc^\dag \hc \rangle \approx 0$. On the other hand, the second row contains local operations on qubit $a$.

The coupling constants in \cref{Eq:CrossResonance} take the form
\begin{align}
  \begin{split}
    \Omega_a &= \uca E_{\text{J},c}'', \Omega_{a;a} = \uca^3 E_{\text{J},c}''' / 2 \\
    \Omega_{a;b} &= \uca \ucb^2 E_{\text{J},c}''', \Omega_{a;c} = \uca \ucc^2 E_{\text{J},c}''',
  \end{split}
\end{align}
where we have defined 
\begin{align}
  \begin{split}
    E_{\text{J},c}'' 
    &=  \alpha  \epsilon   J_1\left(\mu_\alpha \delta\right) \cos \left(\mu_\alpha \bphx \right)  E_{\text{J},c}^{(\alpha)} / \sqrt{2} \\ 
    &+  \beta  J_1\left(\mu_\beta \delta\right) \cos \left(\mu_\beta \bphx \right) 
    E_{\text{J},c}^{(\beta)} / \sqrt{2}, \\
    E_{\text{J},c}''' 
    &= -\alpha  \epsilon  J_1\left(\mu_\alpha \delta\right) \cos \left(\mu_\alpha \bphx \right)  E_{\text{J},c}^{(\alpha)}/\sqrt{2}   \\
    &-\beta   J_1\left(\mu_\beta \delta\right) \cos \left(\mu_\beta \bphx \right) E_{\text{J},c}^{(\beta)}/(\sqrt{2} N^2).
  \end{split}
\end{align}

While in the standard cross-resonance gate protocol the gate is activated by a microwave tone on one of the qubits \cite{rigetti_devoret_2010,chow_et_al_2011}, here it is the coupler mode $c$ that is parametrically driven. This protocol to achieve a CNOT gate is advantageous if the coupler mode is much more strongly coupled to the transmon modes $a$ and $b$ than their direct capacitive coupling. As in the standard cross-resonance protocol \cite{chow_et_al_2011,Sheldon_2016}, the CNOT gate rate $\Omega_{a;b}$ saturates as a function of the amplitude of the parametric drive, in this model due to the Bessel function dependence $J_1\left(\mu_{\alpha,\beta} \delta\right)$. \cref{Tab:Gates} summarizes the different interactions that can be obtained for different choices of modulation frequencies.

\label{Sec:FullCircPT}

\section{Floquet numerics}
\label{Sec:Floq}
In this section we use exact numerical Floquet methods to extract the effective gate Hamiltonian from quasienergy spectra. Floquet theory validates the results obtained using perturbation theory in \cref{Sec:PT,Sec:FullCircPT}. On the other hand, this numerically exact method is applicable beyond the regime of validity of perturbation theory. In this section, we first briefly introduce the method and the notation in \cref{Sec: Floq_meth} and, as an example application, return to our toy model to extract the cross-Kerr interaction $\chi_{ab}$ and the $\sqrt{i\mathrm{SWAP}}$ gate amplitude $J_{ab}$. Using these results, we show how to adjust the system parameters such as to cancel the dynamical cross-Kerr interaction during an $\sqrt{i\mathrm{SWAP}}$ gate. Then, in \cref{Sec:FullNum}, we apply the method to the full circuit Hamiltonian. In particular, we perform a numerical experiment analogous to two-tone spectroscopy for the parametrically driven circuit. For completenes, an introduction to Floquet theory is presented in \cref{Ap:Floquet}.

\subsection{Effective Hamiltonian from Floquet spectra}
\label{Sec: Floq_meth}
Our analysis starts from the observation that the effective Hamiltonian is unitarily equivalent to the Floquet Hamiltonian according to \cref{Eq:HI1,Eq:BCHAna}, and therefore their quasienergy spectra (see \cref{Ap:Floquet}) are identical. In the laboratory frame, we can write
\begin{align}
  \hH_{\textit{eff}} - i\partial_t = e^{-\hG(t)}\left[ \hH(t) - i\partial_t \right]e^{\hG(t)}.
\end{align}
The perturbative expansion for $e^{-\hG(t)}$, and consequently that for $\hH_{\textit{eff}}$, is therefore an iterative approach to finding the Floquet spectrum. 

In this section we compute the Floquet spectrum exactly and show how the parameters of the effective Hamiltonian can be extracted from it. Ac-Stark shifted normal mode frequencies, self- and cross-Kerr interactions, and gate amplitudes are formulated as linear combinations of appropriately identified eigenvalues of the Floquet Hamiltonian. For illustration, in this subsection we will confine our attention to the Floquet analysis of the toy model of \cref{Eq:HTriangle}.

To identify states in the Floquet quasienergy spectrum, we find eigenvectors that have a maximum overlap with a set of known, unperturbed states. We let the state $\ket{i_a i_b i_c}$ be the eigenstate of the time-independent Schr\"odinger equation for the undriven Hamiltonian, that has maximum overlap with the Fock state $\ket{i_a} \ket{i_b} \ket{i_c}$, and denote its eigenenergy by $E_{i_a i_b i_c}$. Finally, we define $\ket{i_a i_b i_c}_F$ as the Floquet eigenmode having maximum overlap with $\ket{i_a i_b i_c}$, and we denote its quasienergy with $\epsilon_{i_a i_b i_c}$. In what follows, we label kets by three integers as above, in the order $a-b-c$.

\begin{figure}[b!]
  \includegraphics[width=1.01\linewidth]{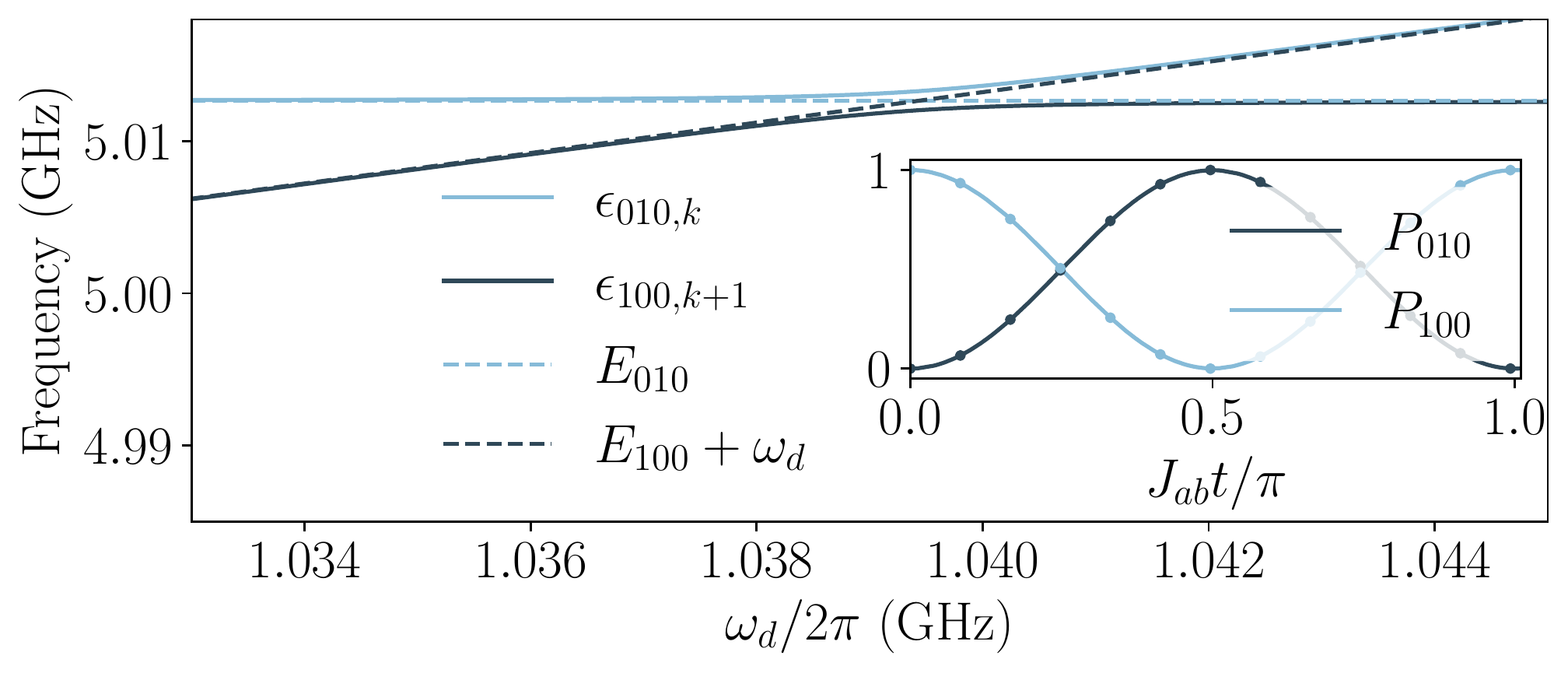}
    \caption{ Quasienergies of the Floquet modes with maximum overlap with eigenstates $|0_a1_b0_c\rangle$ and $|1_a0_b0_c\rangle$, for the toy model. The light and dark blue dashed lines correspond to the eigenenergies of the uncoupled system. The inset shows the population of the Floquet states, $P_{n_a n_b n_c}(t)=|{}_F\langle n_a n_b n_c|\psi(t)\rangle |^2$, compared to the state populations of a two-level system (dots), driven resonantly with Rabi rate $J_{ab}$, where $J_{ab}$ is the gate amplitude obtained from the avoided crossing in the Floquet spectrum. 
    \label{fig:avoidedCrossSim}
  }
  \end{figure}

With these definitions, the gate amplitude $J_{ab}$ has a natural interpretation in the Floquet formalism. As  shown above, the $\sqrt{i\mathrm{SWAP}}$ interaction arises in the toy model if
\begin{align}
  \wdr = \wb - \wa \equiv E_{1 0 0} - E_{0 1 0}. \label{Eq:ResoWdE}
\end{align}
Since the parametric drive enters via a term proportional to $\hbc^\dagger \hbc $, which couples the undriven eigenstates in the two-state manifold $\{\ket{100},\ket{010}\}$, there is an avoided crossing between the Floquet modes $\ket{100;k+1}_F$ and $\ket{010;k}_F$, as shown in \cref{fig:avoidedCrossSim}. Because the gate operation is analogous to Rabi oscillations in the two-state manifold $\{\ket{100},\ket{010}\}$, the size of the avoided crossing is twice the effective gate amplitude, $2 J_{ab}$. For example, if an excitation is originally prepared in the transmon $b$, then population dynamics would obey $P_{010}(t) \equiv | {}_F\langle 010 | \psi(t) \rangle|^2 = \sin^2\left( J_{ab} t \right)$ and $P_{100} = 1 - P_{010}$, in full agreement with exact numerics (inset of \cref{fig:avoidedCrossSim}). Away from the avoided crossing, the difference between the dressed states and the undriven states corresponds to the ac-Stark shift of the transmon normal modes due to the off-resonant drive.

\begin{figure}[t!]
\includegraphics[width=\linewidth]{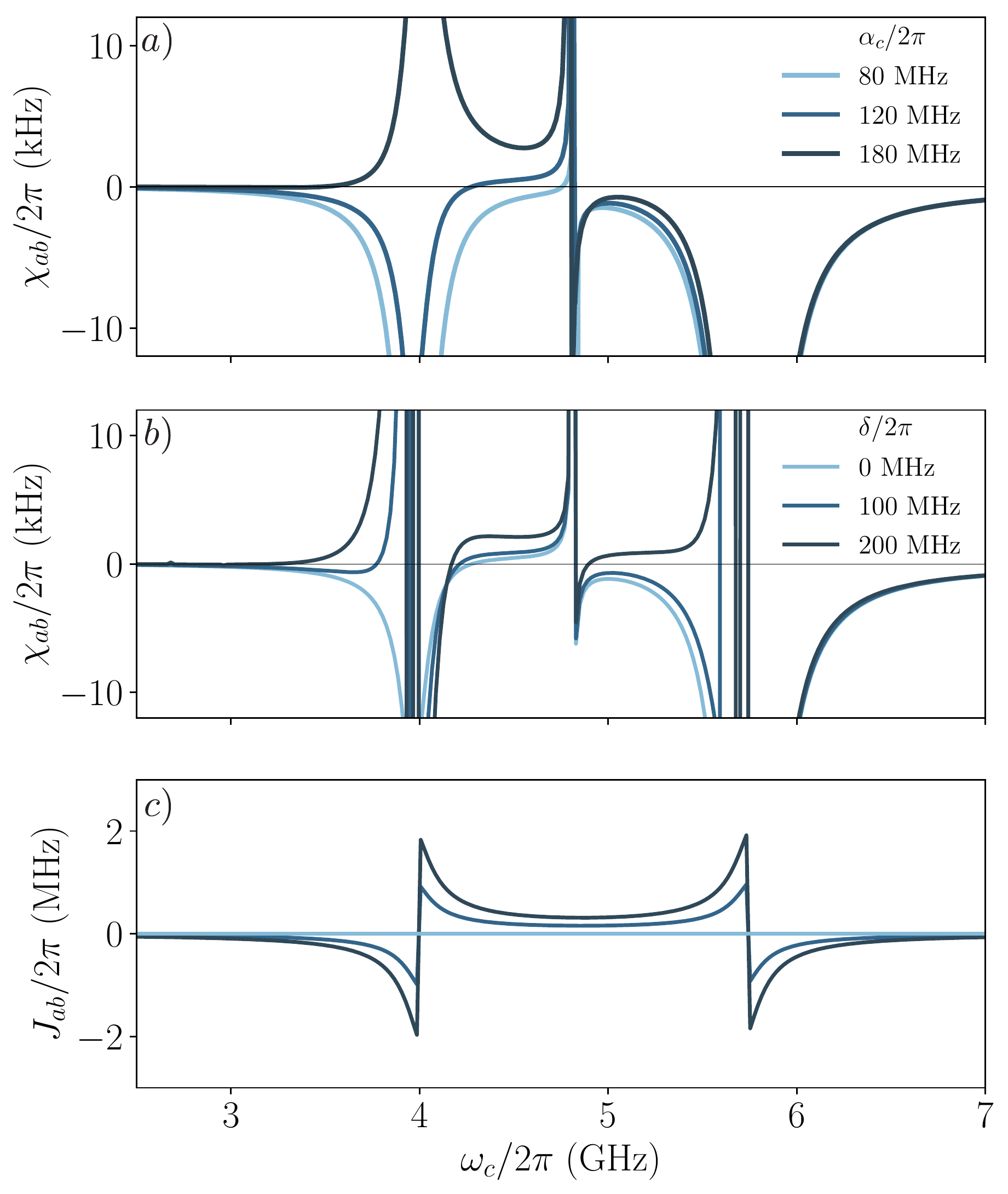}
\caption{a) Static $\ZZ$ interaction at $\delta=0$ for the toy model versus the coupler frequency  for different values of the coupler anharmonicities, with remaining parameters $\Wa/2\pi=4.0$, $\Wb/2\pi=5.75$, $\bala/2\pi=\balb/2\pi=-0.2$, and $\bm{g}_{ac}/2\pi=-\bm{g}_{bc}/2\pi=0.05$ GHz. b) Dynamical $\ZZ$ versus bare coupler frequency for the parameters above and $\balc/2\pi=0.12$ GHz, for different values of the drive amplitude. c) Gate amplitude $J_{ab}$ for the parameters in b).  The Hilbert space dimension for each mode is 5.}
\label{fig:J_JZZ_manyA}
\end{figure}

Note that, in practice, the two-state manifold $\{\ket{100},\ket{010}\}$ is coupled by the drive to other levels. The resonant drive frequency $\wdr$ is then slightly shifted from \cref{Eq:ResoWdE} due to the ac-Stark effect induced by these additional couplings, and the exact value can be determined numerically by minimizing the size of the anticrossing. 

The dynamical cross-Kerr interaction $\chi_{ab}$ is written in terms of a Walsh transform \cite{berke2020transmon} of the quasienergies
\begin{align}
  \chi_{ab}(\delta) = \epsilon_{110} - \epsilon_{100} - \epsilon_{010} + \epsilon_{000},
\end{align}
and reduces to the static cross-Kerr when the parametric drive is turned off:
\begin{align}
\chi_{ab}(0) = E_{110} - E_{100} - E_{010} + E_{000}.
\end{align}
Along with $J_{ab}$ and $\chi_{ab}$, any ac-Stark-shifted quantity pertaining to the effective Hamiltonian can, in principle, be obtained by taking appropriate linear combinations of the quasienergies in the Floquet spectrum. 

Since the Floquet quasienergy spectrum can be obtained from the propagator $\hat{U}(2\pi/\wdr,0)$ over one period of the drive (\cref{Ap:Floquet}), the Floquet method is numerically efficient as compared to the simulation of the dynamics over the complete gate time. The period of the drive is on the order of $1\,\text{ns}$, which is between two and three orders of magnitude shorter than the gate times studied here. Due to its relatively small computational footprint, the Floquet method allows us to efficiently search for optimal gate parameters, \eg a maximal $\J$ with a minimal residual cross-Kerr interaction, $\ZZ$. As an example, in ~\cref{fig:J_JZZ_manyA} we study the behavior of $\J$ and $\ZZ$ as a function of the bare coupler frequency $\Wc$ for different choices of drive amplitude, $\delta$, and bare coupler anharmonicity, $\balc$. 

From these studies we can, for example, find parameters for which the cross-Kerr interaction $\chi_{ab}$ vanishes. As already mentioned, this is important to obtain high-fidelity two-qubit gates and relies on choosing a positive coupler anharmonicity, $\balc$.
In \cref{fig:J_JZZ_manyA} we find that, while varying $\balc$ does not affect $\J$ to lowest order in perturbation theory, it has a considerable impact on $\ZZ$. Indeed, \cref{fig:J_JZZ_manyA}a) shows the static $\ZZ$ for multiple values of $\balc$, and illustrates that it is possible to fine-tune $\balc$ to cancel $\ZZ$. We observe empirically that whenever the bare anharmonicities obey $\bala^{-1}+\balb^{-1}+\balc^{-1} = 0$, one can find $\Wc$ for which $\ZZ=0$.

For the dynamical cross-Kerr interaction $\ZZ$ one observes complex variations with $\delta$. The main resonances appear for $\Wc \approx \Wa$, $\Wb$, and $( \Wa + \Wb )/ 2 $ but the slopes and the sign of $\ZZ$ change and additional resonances appear away from the qubit frequencies, especially when the drive amplitude and coupling strengths $g_{ab,ac}$ are sufficiently large. 
As illustrated in \cref{fig:J_JZZ_manyA}b), by tuning the drive amplitude it is possible to find a bare coupler frequency, $\Wc$, for which the effective $\ZZ(\delta) = 0$. Moreover, the cancellation of $\ZZ(\delta)$ can occur in the form of a `sweet spot' where $\partial \ZZ / \partial \delta \approx 0$. On the other hand, as seen in \cref{fig:J_JZZ_manyA}c), the gate rate increases with $\delta$ without qualitative changes of its dependence on $\Wc$. Therefore, as the gate is turned on or off by varying $\delta$, one can adjust the bare coupler frequency $\Wc$ to maintain the instantaneous $\ZZ(\delta) = 0$. This defines a cross-Kerr-free curve in the parameter space $(\Wc,\delta)$ connecting the `off' point $\delta=0, \ZZ(0)=0, \J(0)=0$ to the `on' point $\delta \neq 0, \ZZ(\delta)=0, \J(\delta) \neq 0$. We study this in detail on the realistic full circuit model in \cref{Sec:FullNum}.

\begin{figure}[t!]
\includegraphics[width=\linewidth]{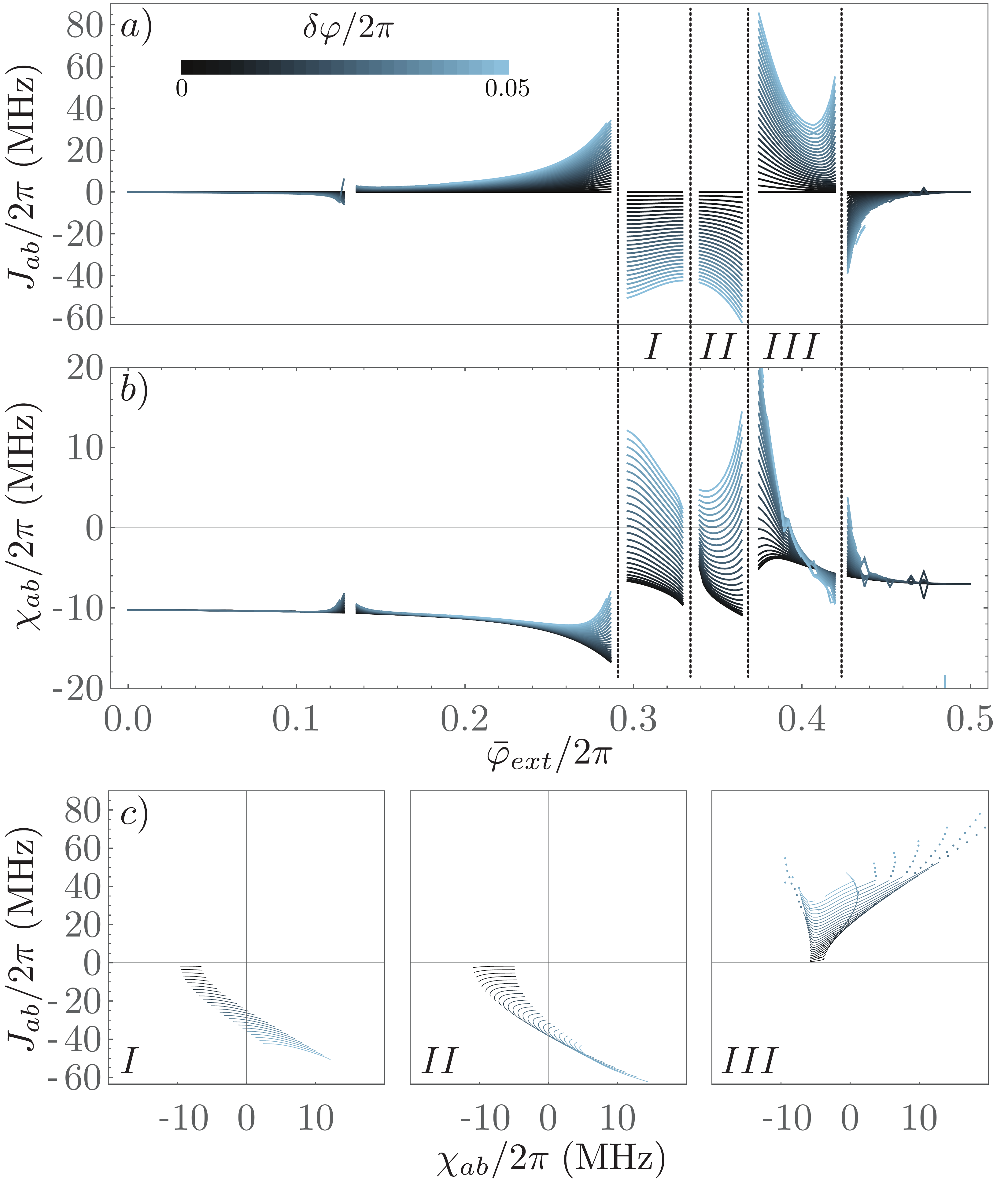}
\caption{Same parameter choices as \cref{Fig:FC_ChiJab_An_Num}. Gate amplitude $\J$ (a) and cross-Kerr $\ZZ$ (b) for different parameteric drive amplitudes $\delta \varphi$ (encoded in curve color) and as a function of the static flux $\bphx$ from Floquet simulations. Data has been excluded where deficient state-tracking in the vicinity of avoided crossings led to unphysical discontinuities in the quantities. c) for regions $I, II$, and $III$ identified in panels a) and b), we eliminate the common parameter $\bphx$ and plot $\J(\ZZ)$. This allows us to identify those regimes in which the $\sqrt{i\mathrm{SWAP}}$ gate interaction can be turned on, while maintaining a vanishing dynamical $\ZZ$.}

\label{fig:fd_J_ZZvsDphi_non-RWA}
\end{figure}

\subsection{Full circuit simulation}
\label{Sec:FullNum}
In this section, we apply the Floquet numerical method to the full circuit Hamiltonian of \cref{Sec:BareModeNumSim}. We study the dependence of the coupling constants in the effective gate Hamiltonian versus DC flux and as a function of the drive amplitude.

\Cref{fig:fd_J_ZZvsDphi_non-RWA} shows the analogues of the plots in \cref{fig:J_JZZ_manyA} for the gate amplitude $\J(\bphx)$ and of the cross-Kerr $\ZZ(\bphx)$ now for the full circuit Hamiltonian. State tracking is performed as described in the previous subsection. However, in the vicinity of avoided crossings, it is impossible to identify with certainty the states generated by the relatively large capacitive couplings considered here. We therefore introduce exclusion regions where state tracking is unreliable. Even though the tracking is expected to be complicated by the presence of counter-rotating terms coupling states with different photon numbers in the full device Hamiltonian of \cref{Sec:BareModeNumSim}, we find that this is not a significant source of tracking error, as compared to errors due to large hybridization.

In \cref{fig:fd_J_ZZvsDphi_non-RWA}b), we represent $\ZZ$ versus the DC-flux $\bphx$ for different values of the flux drive amplitude. Unlike the toy model, $\ZZ$ does not go to zero away from the qubit-coupler resonances. This is because the qubit-coupler detuning saturates as a function of $\bphx$, as opposed to the toy model where the detuning could be increased arbitrarily. In the undriven case (black), we see that for this set of device parameters there does not exist a flux value for which $\ZZ$ vanishes. However, increasing the drive amplitude allows for an active cancellation of the dynamical $\ZZ$ at some flux value. The corresponding behavior of  $\J$ is shown in \cref{fig:fd_J_ZZvsDphi_non-RWA}a). In \cref{fig:fd_J_ZZvsDphi_non-RWA}c), we synthesize the numerical results into three favorable regions of operation for the parametric gate, denoted $I$, $II$, and $III$, respectively [see panel a)]. For these regions, we eliminate the external flux and plot directly the gate amplitude $\J$ against the dynamical cross-Kerr interaction $\ZZ$. This allows us to determine regimes of optimal $\sqrt{i\mathrm{SWAP}}$ gate operation. We conclude that gate amplitudes as high as $40$ MHz, corresponding to a gate time of $12.5$~ns, can be achieved with vanishing cross-Kerr interaction for these parameter choices.

\begin{figure}[t!]
\includegraphics[width=\linewidth]{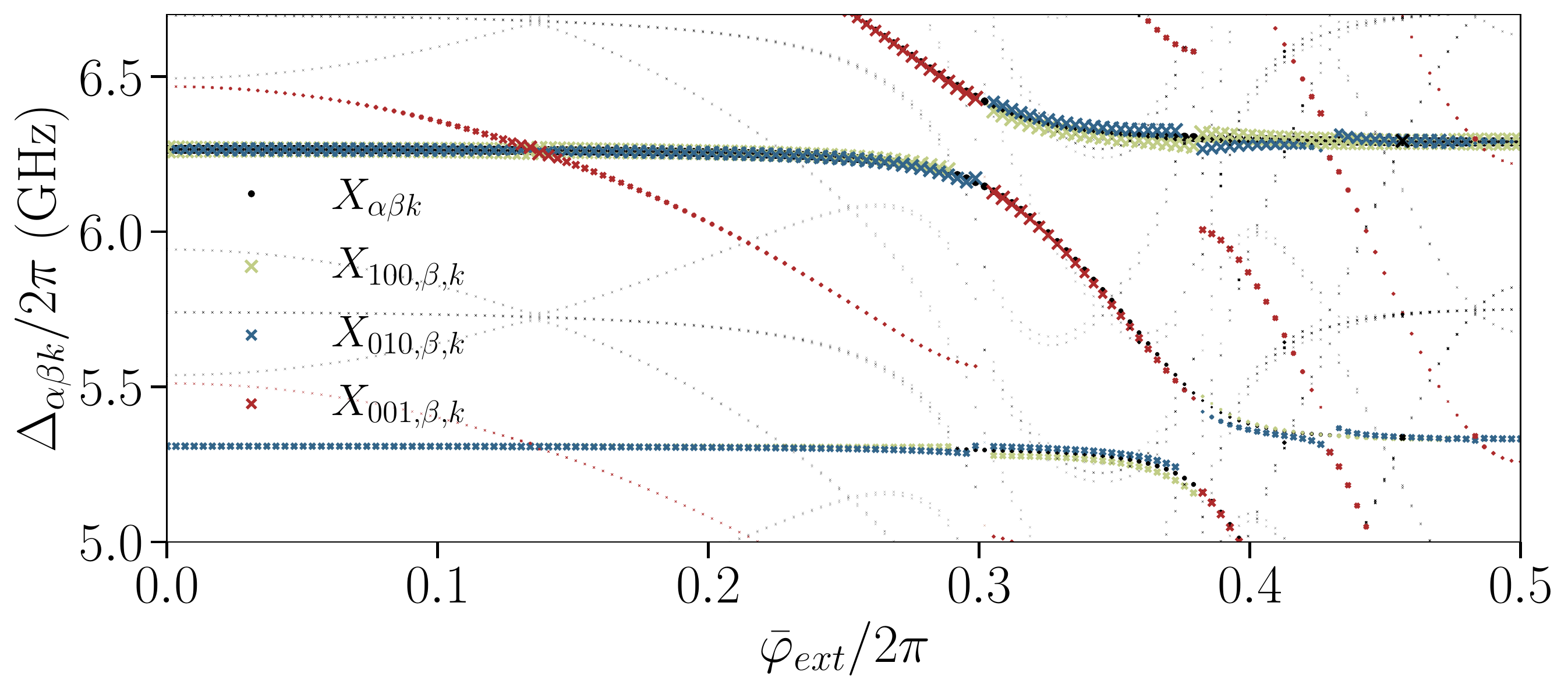}
\caption{Two-tone spectroscopy data from Floquet numerics. Each point corresponds to a possible transition and its size is weighed by the matrix element of the charge operator of bare qubit $a$, $\hat{X}=\bna$. Parameters chosen as in \cref{Fig:FC_ChiJab_An_Num} with $\delta\varphi/2\pi=$ 0.0 (black dots) and 0.03 (crosses). The subscripts $\alpha, \beta$ sweep over the subset of Floquet modes $\{ \ket{000}_F, \ket{100}_F, \ket{010}_F, \ket{001}_F\}$, whereas the drive photon number $k$ takes integer values between $-15$ and $15$.}
\label{fig:spectro_drive_non-RWA}
\end{figure}

For both $\J$ and $\ZZ$ there exist peaks away from the qubit-coupler resonances, situated at $\bar{\varphi}_{ext}/2\pi\approx 0.13, 0.42$. These correspond to avoided crossings appearing in the \textit{driven} Floquet spectrum, corresponding to the hybridization of Floquet levels involving distinct numbers of drive photons. For example, the Floquet level $|100,k\rangle_F$ can couple to the Floquet level $|001,k-1\rangle_F$. This can be seen by unfolding the Floquet spectrum in spectroscopy simulations (see \cref{fig:spectro_drive_non-RWA}).

To exemplify the full extent of the Floquet analysis, we generate two-tone spectroscopy data from our simulations according to \cref{Eq:Dabk,Eq:Xabk} in \cref{Ap:Floquet}, by focusing on the experimentally relevant situation where the parametric drive is on, while the (second) probe tone acts on the bare charge operator $\bna$. In \cref{fig:spectro_drive_non-RWA} we represent the numerically computed spectrum close to the two qubit transition frequencies. The size of each point is proportional to the absolute value of the corresponding matrix element. The black dots correspond to transition frequencies in the undriven spectrum. As expected, the dot sizes are larger for the transitions involving the probed qubit $a$. The large avoided crossings around $\bphx/2\pi \approx \{0.32, 0.38\}$ result from the capacitive couplings between the coupler and the qubits. Secondary avoided crossings appear between the coupler mode and the transmons near $\bar{\varphi}_{ext}/2\pi\approx \{0.13, 0.42\}$ in the driven spectrum, and are responsible for the secondary poles mentioned in the discussion of the coupling constants of the effective Hamiltonian, \cref{fig:fd_J_ZZvsDphi_non-RWA}. Furthermore, as we detail in \cref{Ap:NonRWAEffects}, counterrotating terms induce important corrections when attempting an accurate comparison with spectroscopic data from experiments.

\section{Conclusion}
In summary, we have presented two complementary methods for the analysis of parametrically activated two-qubit gates, one based on analytical time-dependent Schrieffer-Wolff perturbation theory, and one based on numerical Floquet methods. Although we have mostly focused on coupler-mediated parametric $\sqrt{i\mathrm{SWAP}}$ gates, a larger collection of gates can be generated in the same model Hamiltonian. The methods presented here allow one to efficiently evaluate the terms present in the effective gate Hamiltonian.

For the $\sqrt{i\mathrm{SWAP}}$ interaction, we have shown that with experimentally accessible parameters, a gate frequency of $\sim 40$ MHz corresponding to a gate time as short as $12.5$~ns can be obtained with vanishing dynamical cross-Kerr interaction. This fast gate is achieved by working with large capacitive couplings between the qubits and the coupler, while cancelling the cross-Kerr interactions by setting the coupler anharmonicity to positive values, and choosing the right modulation amplitude.
Optimization of realistic device parameters based on close agreements between the Floquet simulations and the experimental data will be published elsewhere \cite{mundada_et_al_2021}.  

We have argued that the analytical method introduced here and which is based on a drive-dependent normal-mode expansion is a computationally efficient strategy to organize the perturbation theory as compared to an energy eigenbasis calculation, for it allows to obtain the parameters of the effective Hamiltonian at lower orders in perturbation theory. Moreover, this strategy is suitable in the regime of comparatively large linear couplings, where the dispersive approximation breaks down. Nonetheless, we have shown that higher orders in analytical perturbation theory are needed for full agreement with exact numerical results, especially for higher-order interactions, such as the dynamical cross-Kerr. Generating higher-order contributions efficiently using computer algebra techniques is the subject of future studies. On the other hand, this work indicates that Floquet numerical methods, as compared to full time-dynamics simulations, is a numerically efficient and exact method for minute optimization studies of parametric gates.

\section*{Acknowledgments}
We thank Joachim Cohen, Moein Malekakhlagh, and Baptiste Royer for useful discussions. We are grateful to Ross Shillito for help with optimizing numerical simulations. This work was undertaken thanks to funding from NSERC, the Canada First Research Excellence Fund, and the U.S. Army Research Office Grant No. W911NF-18-1-0411.

\label{Sec:Conc}

\appendix
\section{Time-dependent Schrieffer-Wolff transformation}
\label{Ap:SWPT}
To obtain equations for $\hGI(t)$, we assume that the generator can be expanded as a series in $\lambda$, that is
\begin{align}
\hGI(t) = \lambda \hGI^{(1)}(t) + \lambda^2 \hGI^{(2)}(t) + \cdots,
\end{align}
and collect powers of $\lambda$ in the BCH expansion of \cref{Eq:BCHAna}
\begin{align}
  \begin{split}
    &e^{-\hGI} (\hI - i\partial_t) e^{\hGI} =  \lambda\hI^{(1)} - i \lambda \dot{\hat{G}}^{(1)}_\text{I}  \\ 
    &+ [\lambda\hI^{(1)},\lambda \hGI^{(1)}] -\frac{i}{2}[\lambda \dot{\hat{G}}^{(1)}_\text{I},\lambda \hGI^{(1)}] - i \lambda^2 \dot{\hat{G}}^{(2)}_\text{I} \\ 
    & - i\partial_t + O(\lambda^3).
  \end{split}
\end{align} 
The above expansion can be expressed compactly
\begin{align}
e^{-\hGI} (\hI - i\partial_t) e^{\hGI} = \lambda^k \sum_{k=1}^{\infty}\left[ \hI^{(k)}(t) - i  \dot{\hat{G}}^{(k)}_\text{I} \right] - i\partial_t.
\end{align}
Provided a prescription for $\lambda^k \hGI^{(k)}(t)$, we have a recursive way of determining higher-order corrections to the interaction Hamiltonian: knowledge of $\lambda \hH_I^{(1)}(t)$ allows one to determine $\lambda^2 \hH_I^{(2)}$, then $\lambda^3 \hH_I^{(3)}$ etc.

The $k^\textit{th}$ order term in the generator, $\lambda^k \hGI^{(k)}(t)$, is determined by the condition that the Hamiltonian be free of oscillatory terms of order $\lambda^k$ or less. This condition can be formulated explicitly if we write, as in \cref{Eq:Decomp1}, 
\begin{align}
  \lambda^k \hH_I^{(k)}(t) = \lambda^k \bHI{k} + \lambda^k \tHI{k}(t) . \label{ApEq:Decompk}
\end{align}
Then oscillatory terms $\lambda^k \tHI{k}$ are canceled for every $k$ if
\begin{align}\label{ApEq:RWACondition}
   \lambda^k \hGI^{(k)}(t) = \frac{1}{i} \int_0^t \lambda^k \tHI{k}(t). 
\end{align} 
Note that, in the above expression, we have imposed the boundary condition $\hGI^{(k)}(0)=0$ by specifying the lower limit of the integration. Noting that \cref{ApEq:RWACondition} implies
\begin{align}
\begin{split}
  \lambda^k \widetilde{\hGI}^{(k)}(t) =& \frac{1}{i} \int^t \lambda^k \tHI{k}(t), \\
  \lambda^k \overline{\hGI}^{(k)}(t) =& \frac{1}{i} \left[ \int^t \lambda^k \tHI{k}(t) \right]_{t=0}.
\end{split}
\end{align} 
The DC part of the generator, $\lambda^k \overline{\hGI}^{(k)}$, is non-vanishing here as a result of the boundary condition in \cref{ApEq:RWACondition}, as opposed to the zero time-average property of kick operators, to which the generator studied here is related \cite{goldman_dalibard_2014}. 

With the above formalism in place, we are now ready to compute perturbative corrections. From \cref{Eq:BCHAna} we identify the $\lambda^2$ correction to the interaction-picture Hamiltonian
\begin{align}
\lambda^2 \hI^{(2)}(t) = [\lambda\hI^{(1)},\lambda \hGI^{(1)}] -\frac{i}{2}[\lambda \dot{\hat{G}}^{(1)}_\text{I},\lambda \hGI^{(1)}].
\end{align}
Going ahead and solving the RWA condition in \cref{ApEq:RWACondition} at order $\lambda^1$, we find the order-$\lambda^2$ RWA Hamiltonian
\begin{align}
  \begin{split}
  \lambda^2\bHI{2} = & \frac{1}{i} \overline{\left[\bHI{1}, \int_0^t  \lambda \tHI{1}(t') dt'\right]}  \\ & + \frac{1}{2i}\overline{ \left[  \lambda\tHI{1}(t), \int_0^t  \lambda \tHI{1}(t') dt' \right] }. \label{ApEq:bHI2}
  \end{split}
\end{align}
This procedure can be iterated to higher orders, with increasing complexity due to the proliferation of terms from nested commutators in the BCH expansion.

\section{Circuit quantization}
\label{App:Quantization}

\begin{figure}
  \includegraphics[width=\linewidth]{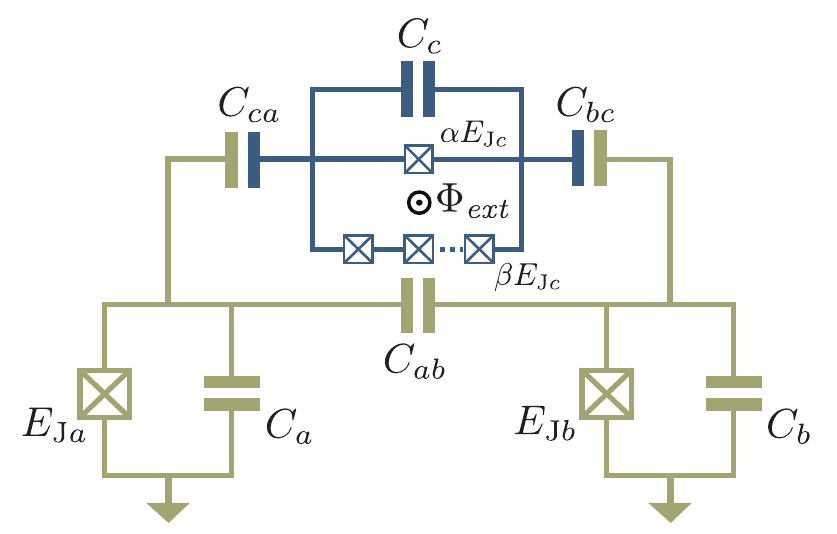}
  \caption{Circuit schematic and notations used in the derivation  of the circuit Lagrangian in \cref{App:Quantization}. The coupler consists of two branches of total capacitances $C_\alpha$ and $C_\beta$ (not indicated in the figure). The $\alpha$ branch consists of a single Josephson junction, while the `$\beta$' branch $N$ junctions in series. The bare coupler and transmon modes are connected capacitively through coupling capacitances $C_{ab,bc,ca}$.\label{Fig:CircuitCapacitances}}
\end{figure}

In this appendix we derive the model Hamiltonian of \cref{Eq:Model} from the circuit Lagrangian corresponding to \cref{Fig:CircuitCapacitances}. Assuming the individual modes of the junction array have small impedance, guaranteed by sufficiently large Josephson energy, the junction array can be described by an effective one-dimensional Lagrangian where the total phase difference across the array is spread evenly through the junctions. The effective one-dimensional Lagrangian associated with the bare coupler mode is 
\begin{align}
  \mathcal{L}_c = \sum_{k=\alpha,\beta} \frac{C_k}{2}\dot{\boldsymbol{\phi}}_k^2 + \alpha \EJc \cos\left[\boldsymbol{ \varphi}_\alpha\right] + \beta N  \EJc \cos\left[ \frac{\boldsymbol{\varphi}_\beta}{N}\right], 
\end{align}
where $\boldsymbol{\phi}_\alpha$ is the branch flux across the small junction and the shunt capacitor with total capacitance $C_\alpha$,  $\boldsymbol{\phi}_\beta$ is the branch flux across the junction array with effective capacitance $C_\beta$, and  $\boldsymbol{\varphi}_k =  2\pi \boldsymbol{\phi}_k /\Phi_0$ are the associated reduced phase variables, and $\Phi_0$ is the superconducting flux quantum. The phases $\boldsymbol{\varphi}_\alpha$ and $\boldsymbol{\varphi}_\beta$ are constrained by the fluxoid quantization, $\boldsymbol{\varphi}_\alpha + \boldsymbol{\varphi}_\beta = \varphi_{\rm ext}$. We define the new coordinates 
\begin{align}
\begin{split}
    \boldsymbol{\varphi}_\alpha & =   \boldsymbol{\varphi}_c + \mu_\alpha \varphi_{\rm ext}, \\
    \boldsymbol{\varphi}_\beta & =   -\boldsymbol{\varphi}_c - N \mu_\beta \varphi_{\rm ext}, 
    \end{split}
\end{align}
with $\mu_\alpha - N \mu_\beta = 1$, such that the capacitive energy in the Lagrangian is now purely quadratic in $\dot{\boldsymbol{\phi}}_c$. We thus require $C_\alpha \mu_\alpha + C_\beta N \mu_\beta = 0$. We obtain 
\begin{align}
\begin{split}
   \mu_\alpha  & =  \frac{C_\beta}{C_\alpha + C_\alpha}, \\
   \mu_\beta & =  -\frac{1}{N}\frac{C_\alpha}{C_\alpha + C_\beta}.
   \end{split}
\end{align}

Up to time-dependent scalar terms, we obtain the form
\begin{align}
\begin{split}
  \mathcal{L}_c =  \frac{C_c}{2}\dot{\boldsymbol{\phi}}_c^2 + \alpha \EJc \cos\left[\boldsymbol{\varphi}_c + \mu_\alpha \varphi_{\rm ext}\right] \\
   + \beta N  \EJc \cos\left[ \frac{\boldsymbol{\varphi}_c}{N} + \mu_\beta \varphi_{\rm ext}\right]. 
  \end{split}
\end{align}
Moreover, for the two bare transmon modes $j=a,b$ the Lagrangian reads
\begin{align}
  \mathcal{L}_j =  \frac{C_j}{2}\dot{\boldsymbol{\phi}}_j^2 + E_{\text{J}j} \cos  \boldsymbol{\varphi}_j.
\end{align}

The total Lagrangian of the system then takes the form 
\begin{align}
  \mathcal{L} =  \mathcal{L}_a + \mathcal{L}_b + \mathcal{L}_c + \mathcal{L}_g,
\end{align}
where we have introduced the capacitive coupling between the three bare modes 
\begin{align}
  \mathcal{L}_g = \frac{C_{ab}}{2} \dot{\boldsymbol{\phi}}_a \dot{\boldsymbol{\phi}}_b + \frac{C_{bc}}{2} \dot{\boldsymbol{\phi}}_b \dot{\boldsymbol{\phi}}_c + \frac{C_{ca}}{2} \dot{\boldsymbol{\phi}}_c \dot{\boldsymbol{\phi}}_a.
\end{align}

\section{Perturbation theory for the toy model}
\label{Ap:PTToy}
In this section, we reproduce expressions for the cross-Kerr interaction obtained to second-order in perturbation theory for the toy model. The full expression of the second-order RWA correction to the cross-Kerr interaction in \cref{Subsec:EffGateToy} reads
\begin{align} \label{ApEq:Chi2RWAToy}
\begin{split}
  \chi_{ab,\text{\cref{Subsec:EffGateToy}}}^{(2)} &=  \frac{4 \left(\sum_{j=a,b,c}\uaj^2 \ubj \ucj \balj \right)^2}{\wb-\wc}  \\
  &+ \frac{4 \left(\sum_{j=a,b,c} \uaj \ubj^2 \ucj \balj\right)^2}{\wa-\wc}  \\
&+ \frac{2 \left(\sum_{j=a,b,c}  \uaj \ubj \ucj^2 \balj \right)^2}{\wa+\wb-2 \wc}  \\
&- \frac{2 \left(\sum_{j=a,b,c} \uaj^3 \ubj \balj  \right)^2}{\wa-\wb} \\
&+ \frac{2\left(\sum_{j=a,b,c} \uaj \ubj^3 \balj \right)^2}{\wa-\wb} \\
&+ \frac{\uac \ubc \sum_{j=a,b,c} \uaj \ubj  \left(\uaj^2-\ubj^2\right)  }{\wa-\wb} \delta.
\end{split}
\end{align}

The second-order correction to the static cross-Kerr interaction as calculated in \cref{Sec:AnhCorrections} is:
\begin{align}\label{ApEq:Chi2RWAToyMat}
\begin{split}
  \chi_{ab,\text{\cref{Sec:AnhCorrections}}}^{(2)} &=  \frac{4 \left(\sum_{j=a,b,c} \uaj^2 \ubj \ucj \balj\right)^2}{\omega_b-\omega_c + \sum_{j=a,b,c} 2 \uaj^2  \left(\ubj^2-\ucj^2\right) \balj  } \\
  &+\frac{4 \left(\sum_{j=a,b,c} \uaj \ubj^2 \ucj \balj \right)^2}{\omega_a-\omega_c+ \sum_{j=a,b,c} 2 \ubj^2  \left(\uaj^2-\ucj^2\right)\balj} \\
  &+\frac{2 \left(\sum_{j=a,b,c} \uaj \ubj \ucj^2 \balj\right)^2}{\omega_a+\omega_b-2 \omega_c+\left(2 \uaj^2 \ubj^2-\ucj^4\right)\balj } \\
  &-\frac{2 \left(\sum_{j=a,b,c}\uaj^3 \ubj \balj\right)^2}{\omega_a-\omega_b+\sum_{j=a,b,c} \left(\uaj^4-2 \uaj^2 \ubj^2\right) \balj } \\
  &+\frac{2 \left(\sum_{j=a,b,c} \uaj \ubj^3 \balj \right)^2}{\omega_a-\omega_b+\sum_{j=a,b,c} \left(2 \uaj^2 \ubj^2-\ubj^4\right) \bala }. 
\end{split}
\end{align}
The expression for the dynamical cross-Kerr interaction, $\chi_{ab,\text{\cref{Sec:AnhCorrections}}}^{(2)}$ at $\delta \neq 0$, is available from the formalism, but it is too lengthy to be reproduced here. In the main text, an evaluation of this expression is used in making direct comparisons to exact numerics.

\section{Details for full circuit Hamiltonian}
In this Appendix, we record a number of results used in \cref{Sec:CircNOJA}, in particular the formulae used for normal-ordered expansions in \cref{Ap:NOExp}, and the time-dependent terms in the coupler Hamiltonian in \cref{Ap:TDCplH}. 
\label{Ap:FullCircuitTheory}

\subsection{Normal-ordered expansions of trigonometric functions. Jacobi-Anger expansions}
\label{Ap:NOExp}

Sine and cosine are expanded in normal order using the following two expressions \cite{Marcos_2013} [recall that $\pha = \sqrt{\eta_a/2} (\ha+\ha^\dagger)$]
\begin{align}
\begin{split}
  \cos \pha  &=  e^{-\frac{\eta_a}{4}} \sum\limits_{\substack{m,n \geq 0\\m+n = \text{ even}}} \frac{\left( -\frac{\eta_a}{2}\right)^\frac{m+n}{2}\ha^{\dagger m} \ha^n} {m! n!}  , \\
  \sin \pha  &= e^{-\frac{\eta_a}{4}} \sqrt\frac{\eta_a}{2}  \sum\limits_{\substack{m,n \geq 0\\m+n = \text{ odd}}} \frac{\left( -\frac{\eta_a}{2}\right)^\frac{m+n-1}{2}\ha^{\dagger m} \ha^n} {m! n!} ,     \label{Eq:CosSinNO}
   \end{split}
\end{align}
with analogous expressions for the operators $\hb$ and $\hc$. 

\subsection{Time-dependent terms in the coupler Hamiltonian}
\label{Ap:TDCplH}
Terms corresponding to the Jacobi-Anger expansion up to the second harmonic of the drive in the bare coupler Hamiltonian $\widetilde{\hH}_c(t)$ in \cref{Sec:CircNOJA} are listed in \cref{Tab:TildeHcQuartic}. The operator monomial at the beginning of each row is to be multiplied by the sum of the two following columns, and then results from all rows are to be summed. The coefficients of the missing monomials $\hbc, \hbc^2, \hbc^3, \hbc^\dagger \hbc^2, \hbc^4, \hbc^\dagger \hbc^3$ are obtained by Hermitian conjugation.

\begin{table*}
\begin{tabular}{ccc}
  Monomial & $J_1(\delta )$ & $J_2(\delta )$ \\
  \hline \\
 $\hbc^\dagger$ & $\sqrt{2} \alpha  \epsilon  e^{-\frac{\eta_c}{4}} \sqrt{\eta_c}
   \EJc J_1\left(\delta  \mu_\alpha\right) \sin \left(t \omega_d\right) \cos \left(\mu
  _\alpha \bphx\right)+$ & $\sqrt{2} \alpha  \epsilon  e^{-\frac{\eta_c}{4}} \sqrt{\eta_c}
   \EJc J_2\left(\delta  \mu_\alpha\right) \cos \left(2 t \omega_d\right) \sin \left(\mu
  _\alpha \bphx\right)+$ \\
  $\,$ & $\sqrt{2} \beta  N e^{-\frac{\eta_c}{4 N^2}} \sqrt{\frac{\eta_c}{N^2}}
    \EJc J_1\left(\delta  \mu_\beta\right) \sin \left(t \omega_d\right) \cos \left(\mu
   _\beta \bphx\right)$ & $\sqrt{2} \beta  N e^{-\frac{\eta_c}{4 N^2}} \sqrt{\frac{\eta
   _c}{N^2}} \EJc J_2\left(\delta  \mu_\beta\right) \cos \left(2 t
    \omega_d\right) \sin \left(\mu_\beta \bphx\right)$ \\
  $\hbc^\dagger \hbc^\dagger$ & $-\frac{1}{2} \alpha  \epsilon  e^{-\frac{\eta_c}{4}} \eta_c
    \EJc J_1\left(\delta  \mu_\alpha\right) \sin \left(t \omega_d\right) \sin \left(\mu
   _\alpha \bphx\right)$ & $\frac{1}{2} \alpha  \epsilon  e^{-\frac{\eta_c}{4}} \eta_c
    \EJc J_2\left(\delta  \mu_\alpha\right) \cos \left(2 t \omega_d\right) \cos \left(\mu
   _\alpha \bphx\right)$ \\
  \, & $-\frac{\beta  \eta_c e^{-\frac{\eta_c}{4 N^2}} \EJc
    J_1\left(\delta  \mu_\beta\right) \sin \left(t \omega_d\right) \sin \left(\mu_\beta \bphx\right)}{2
    N}$ & $+\frac{\beta  \eta_c e^{-\frac{\eta_c}{4 N^2}} \EJc
    J_2\left(\delta  \mu_\beta\right) \cos \left(2 t \omega_d\right) \cos \left(\mu_\beta \bphx\right)}{2
    N}$ \\
 $\hbc^\dagger \hbc$ & $-\alpha  \epsilon  e^{-\frac{\eta_c}{4}} \eta_c
   \EJc J_1\left(\delta  \mu_\alpha\right) \sin \left(t \omega_d\right) \sin \left(\mu
  _\alpha \bphx\right)$ & $+\alpha  \epsilon  e^{-\frac{\eta_c}{4}} \eta_c
   \EJc J_2\left(\delta  \mu_\alpha\right) \cos \left(2 t \omega_d\right) \cos \left(\mu
  _\alpha \bphx\right)$ \\
  $\,$ & $-\frac{\beta  \eta_c e^{-\frac{\eta_c}{4 N^2}}
   \EJc J_1\left(\delta  \mu_\beta\right) \sin \left(t \omega_d\right) \sin \left(\mu
  _\beta \bphx\right)}{N}$ & $+\frac{\beta  \eta_c e^{-\frac{\eta_c}{4 N^2}}
   \EJc J_2\left(\delta  \mu_\beta\right) \cos \left(2 t \omega_d\right) \cos \left(\mu
  _\beta \bphx\right)}{N}$ \\
 $\hbc^\dagger \hbc^\dagger \hbc^\dagger$ & $-\frac{\alpha  \epsilon  e^{-\frac{\eta_c}{4}} \eta_c^{3/2}
   \EJc J_1\left(\delta  \mu_\alpha\right) \sin \left(t \omega_d\right) \cos \left(\mu
  _\alpha \bphx\right)}{6 \sqrt{2}}$ & $-\frac{\alpha  \epsilon  e^{-\frac{\eta_c}{4}} \eta_c^{3/2}
   \EJc J_2\left(\delta  \mu_\alpha\right) \cos \left(2 t \omega_d\right) \sin \left(\mu
  _\alpha \bphx\right)}{6 \sqrt{2}}$ \\
 \, & $-\frac{\beta  \eta_c e^{-\frac{\eta_c}{4 N^2}} \sqrt{\frac{\eta_c}{N^2}}
   \EJc J_1\left(\delta  \mu_\beta\right) \sin \left(t \omega_d\right) \cos \left(\mu
  _\beta \bphx\right)}{6 \sqrt{2} N}$ & $-\frac{\beta  \eta_c e^{-\frac{\eta_c}{4
   N^2}} \sqrt{\frac{\eta_c}{N^2}} \EJc J_2\left(\delta  \mu_{\beta
   }\right) \cos \left(2 t \omega_d\right) \sin \left(\mu_\beta \bphx\right)}{6 \sqrt{2} N}$ \\
 $\hbc^\dagger \hbc^\dagger \hbc$ & $-\frac{\alpha  \epsilon  e^{-\frac{\eta_c}{4}} \eta_c^{3/2}
   \EJc J_1\left(\delta  \mu_\alpha\right) \sin \left(t \omega_d\right) \cos \left(\mu
  _\alpha \bphx\right)}{2 \sqrt{2}}$ & $-\frac{\alpha  \epsilon  e^{-\frac{\eta_c}{4}} \eta_c^{3/2}
   \EJc J_2\left(\delta  \mu_\alpha\right) \cos \left(2 t \omega_d\right) \sin \left(\mu
  _\alpha \bphx\right)}{2 \sqrt{2}}$ \\
 $\,$ & $-\frac{\beta  \eta_c e^{-\frac{\eta_c}{4 N^2}} \sqrt{\frac{\eta_c}{N^2}}
   \EJc J_1\left(\delta  \mu_\beta\right) \sin \left(t \omega_d\right) \cos \left(\mu
  _\beta \bphx\right)}{2 \sqrt{2} N}$ & $-\frac{\beta  \eta_c e^{-\frac{\eta_c}{4
   N^2}} \sqrt{\frac{\eta_c}{N^2}} \EJc J_2\left(\delta  \mu_{\beta
   }\right) \cos \left(2 t \omega_d\right) \sin \left(\mu_\beta \bphx\right)}{2 \sqrt{2} N}$ \\
 $\hbc^\dagger \hbc^\dagger \hbc^\dagger \hbc^\dagger$ & $\frac{1}{48} \alpha  \epsilon  e^{-\frac{\eta_c}{4}}
   \eta_c^2 \EJc J_1\left(\delta  \mu_\alpha\right) \sin \left(t \omega
  _d\right) \sin \left(\mu_\alpha \bphx\right)$ & $-\frac{1}{48} \alpha  \epsilon  e^{-\frac{\eta_c}{4}} \eta
  _c^2 \EJc J_2\left(\delta  \mu_\alpha\right) \cos \left(2 t \omega_d\right)
   \cos \left(\mu_\alpha \bphx\right)$ \\
 \, & $+\frac{\beta  \eta_c^2 e^{-\frac{\eta_c}{4 N^2}} \EJc
   J_1\left(\delta  \mu_\beta\right) \sin \left(t \omega_d\right) \sin \left(\mu_\beta \bphx\right)}{48
   N^3}$ & $-\frac{\beta  \eta_c^2 e^{-\frac{\eta_c}{4 N^2}}
   \EJc J_2\left(\delta  \mu_\beta\right) \cos \left(2 t \omega_d\right) \cos \left(\mu
  _\beta \bphx\right)}{48 N^3}$ \\
 $\hbc^\dagger \hbc^\dagger \hbc^\dagger \hbc$ & $\frac{1}{12} \alpha  \epsilon  e^{-\frac{\eta_c}{4}}
   \eta_c^2 \EJc J_1\left(\delta  \mu_\alpha\right) \sin \left(t \omega
  _d\right) \sin \left(\mu_\alpha \bphx\right)$ & $-\frac{1}{12} \alpha  \epsilon  e^{-\frac{\eta_c}{4}} \eta
  _c^2 \EJc J_2\left(\delta  \mu_\alpha\right) \cos \left(2 t \omega_d\right)
   \cos \left(\mu_\alpha \bphx\right)$ \\
 \,  & $+\frac{\beta  \eta_c^2 e^{-\frac{\eta_c}{4 N^2}} \EJc
   J_1\left(\delta  \mu_\beta\right) \sin \left(t \omega_d\right) \sin \left(\mu_\beta \bphx\right)}{12
   N^3}$ & $-\frac{\beta  \eta_c^2 e^{-\frac{\eta_c}{4 N^2}}
   \EJc J_2\left(\delta  \mu_\beta\right) \cos \left(2 t \omega_d\right) \cos \left(\mu
  _\beta \bphx\right)}{12 N^3}$ \\
 $\hbc^\dagger \hbc^\dagger \hbc \hbc$ & $+\frac{1}{8} \alpha  \epsilon  e^{-\frac{\eta_c}{4}} \eta
  _c^2 \EJc J_1\left(\delta  \mu_\alpha\right) \sin \left(t \omega_d\right)
   \sin \left(\mu_\alpha \bphx\right)$ & $-\frac{1}{8} \alpha  \epsilon  e^{-\frac{\eta_c}{4}} \eta
  _c^2 \EJc J_2\left(\delta  \mu_\alpha\right) \cos \left(2 t \omega_d\right)
   \cos \left(\mu_\alpha \bphx\right)$ \\
 \; & $+\frac{\beta  \eta_c^2 e^{-\frac{\eta_c}{4
   N^2}} \EJc J_1\left(\delta  \mu_\beta\right) \sin \left(t \omega_d\right)
   \sin \left(\mu_\beta \bphx\right)}{8 N^3}$ & $-\frac{\beta  \eta_c^2 e^{-\frac{\eta_c}{4 N^2}}
   \EJc J_2\left(\delta  \mu_\beta\right) \cos \left(2 t \omega_d\right) \cos \left(\mu
  _\beta \bphx\right)}{8 N^3}$ 
 \end{tabular}
\caption{\label{Tab:TildeHcQuartic} Time-dependent terms, up to quartics, in the bare coupler Hamiltonian.}
\end{table*}

\section{Normal-mode transformation}
\label{App:NormalMode}

In \cref{Sec:CircNOJA}, we made use of a normal mode transformation that eliminates the off-diagonal capacitive coupling terms from the time-independent quadratic Hamiltonian. In this section we provide the steps to obtain the normal mode coefficients.

Consider the quadratic form (repeated indices are summed over):
\newcommand{\bnal}{\hat{\bar{n}}_\alpha}
\newcommand{\bnbe}{\hat{\bar{n}}_\beta}
\newcommand{\bpal}{\hat{\bar{\varphi}}_\alpha}
\newcommand{\bpbe}{\hat{\bar{\varphi}}_\beta}
\begin{align}
  \hH = A_{\alpha\beta} \bnal \bnbe + B_{\alpha \beta} \bpal \bpbe.
\end{align}
We make a simplification by assuming that there are no off-diagonal inductive terms, $B_{\alpha \beta} \propto \delta_{\alpha\beta}$, which is valid for the circuit studied here. The diagonalization involves three steps:

\textit{Step 1.} Rescale the variables so that the diagonal part of the Hamiltonian, the inductive part, contains terms with the \textit{ same } inductive energy. For this, let us define the square root of the product of the inductive energies $B = \left(\prod_\alpha B_{\alpha\alpha}\right)^{1/2}$ and the dimensionless coefficients $f_\alpha = \sqrt{B/B_{\alpha\alpha}}$. Then we introduce new canonically-conjugate coordinates:
\begin{align}
  \hat{\varphi}_\alpha' = f_\alpha^{-1} \bpal, \qquad \hat{n}_\alpha' = f_\alpha \bnal .
\end{align}
In terms of the new coordinates, and letting $A_{\alpha\beta}' = A_{\alpha\beta}/ (f_\alpha f_\beta)$ (no implicit summation), we have
\begin{align}
  \hH = A'_{\alpha \beta} \hat{n}_\alpha' \hat{n}_\beta' + B\delta_{\alpha\beta} \hat{\varphi}_\alpha' \hat{\varphi}_\beta'.
\end{align}

\textit{Step 2.} Diagonalize the capacitive coupling matrix $A'$. We assume here that this is possible and is achieved by an orthonormal matrix $S$, such that 
\begin{align}
  A'_{\alpha \beta} = (S^T)_{\alpha \mu} D_{\mu \nu} S_{\nu \beta} = S_{\mu \alpha} D_{\mu \nu} S_{\nu \beta},
\end{align}  
with $D_{\mu \nu}$ a diagonal matrix. Rewriting the above as $A'_{\alpha \beta} (S^T)_{\beta \gamma} = (S^T)_{\alpha \mu} D_{\mu \nu} S_{\nu \beta} (S^T)_{\beta \gamma} = (S^T)_{\alpha \mu} D_{\mu \gamma}$, or $\mathbf{A}'\cdot \mathbf{S}^T = \mathbf{S}^T \cdot \mathbf{D}$, then the matrix $\mathbf{S}$ contains the eigenvectors of $\mathbf{A}'$ on its \textit{rows}.  This diagonalization leads to
\begin{align}
  \hH = S_{\mu \alpha} D_{\mu \nu} S_{\nu \beta} \hat{n}_\alpha' \hat{n}_\beta' + B\delta_{\alpha\beta} \hat{\varphi}_\alpha' \hat{\varphi}_\beta'.
\end{align}
Inspecting the first term, we again define new coordinates
\begin{align}
  \hat{n}_\mu'' = S_{\mu \alpha} \hat{n}_\alpha',\qquad \hat{\varphi}_\mu'' = S_{\mu \alpha} \hat{\varphi}_\alpha'.
\end{align}
One can verify that the new double-primed coordinates are canonically conjugate because the transformation is orthonormal: $[\hat{n}_\mu'',\hat{\varphi}_\nu''] = S_{\mu \alpha}S_{\nu \beta} [ \hat{n}_\alpha',  \hat{\varphi}_\beta' ] = i S_{\mu \alpha}S_{\nu \beta} \delta_{\alpha\beta}  = i S_{\mu \alpha}S_{\nu \alpha}  = i \delta_{\mu \nu}$. With this, we obtain a diagonal form for the Hamiltonian
\begin{align}
  \hH = \hat{n}_\alpha'' D_{\alpha \beta} \hat{n}_\beta'' + B\delta_{\alpha\beta} \hat{\varphi}_\alpha'' \hat{\varphi}_\beta'',
\end{align}
where in the second term we used the fact that the orthogonal transformation preserves the inner product. 

\textit{Step 3.} Finally, we need to undo the rescaling transformation of Step 1. That is, introduce a third and last pair of canonically conjugate coordinates, the \textit{normal-mode coordinates}
\begin{align}
  \hat{\varphi}_\alpha = f_\alpha \hat{\varphi}_\alpha'',\qquad \hat{n}_\alpha =  f_\alpha^{-1} \hat{n}_\alpha''.
\end{align}
At last the quadratic Hamiltonian reads
\begin{align}
  \begin{split}
    \hH &= \hat{n}_\alpha  f_\alpha f_\beta D_{\alpha \beta} \hat{n}_\beta  + \frac{B\delta_{\alpha\beta}}{f_\alpha f_\beta} \hat{\varphi}_\alpha \hat{\varphi}_\beta \\
    &= \hat{n}_\alpha  f_\alpha f_\beta D_{\alpha \beta} \hat{n}_\beta  +  \hat{\varphi}_\alpha B_{\alpha\beta}  \hat{\varphi}_\beta.
  \end{split}
\end{align}
This is the final normal-mode Hamiltonian. 

\textit{Hybridization coefficients.} It is helpful to summarize the normal mode transformation by skipping over the intermediate variables (primed, and double-primed). For this we have to invert the definitions of the intermediate coordinates to obtain
\begin{align}
  \begin{split}
    \bpal &=  \sum_\beta f_\alpha S_{\beta\alpha} f_\beta^{-1} \hat{\varphi}_\beta \equiv \sum_\beta U_{\alpha \beta} \hat{\varphi}_\beta, \\
  \bnal &= \sum_\beta f_\alpha^{-1} S_{\beta\alpha} f_\beta \hat{n}_\beta \equiv \sum_\beta V_{\alpha \beta} \hat{n}_\beta, 
  \end{split}
\end{align}
where we have used $\hat{\varphi}'_\alpha = S_{\mu \alpha} \hat{\varphi}_\mu''$ and $\hat{n}'_\alpha = S_{\mu \alpha} \hat{n}_\mu''$. Note that $\mathbf{U}\cdot\mathbf{V}^T=1$, \ie the transformation from bare to normal modes is canonical.

\textit{Creation and annihilation operators.} Lastly, we consider the creation and annihiliation operators. In order for squeezing terms to disappear in the Hamiltonian, we need:
\begin{align}\label{ApEq:NMTransf}
  \begin{split}
    \hat{\bar{\varphi}}_\alpha &= \sum_{\beta=a,b,c} \frac{u_{\alpha\beta}}{\sqrt{2}} (\hat{\beta}+\hat{\beta}^\dagger), \\
    \hat{\bar{n}}_\alpha &= \sum_{\beta=a,b,c} \frac{v_{\alpha\beta}}{i\sqrt{2}} (\hat{\beta}-\hat{\beta}^\dagger), 
  \end{split}
\end{align}
where 
\begin{align}
  u_{\alpha\beta} = U_{\alpha \beta} \sqrt{\epsilon_\beta},\;  v_{\alpha \beta} = \frac{V_{\alpha \beta}}{\sqrt{\epsilon_\beta}}.
\end{align}
Finally, we have obtained the hybridization coefficients entering \cref{Eq:NMTransf} in the main text. The approach given in this Appendix generalizes to an arbitrary number of modes with off-diagonal coupling in either the capacitive matrix, or in the inductive matrix. 

\section{Floquet theory}
\label{Ap:Floquet}
This appendix provides a practical summary of Floquet theory. The spectrum of a monochromatically driven system can be obtained from the Floquet formalism~\cite{grifoni_haenggi_1998}, according to which the time-dependent Schr\"odinger equation for a periodically driven Hamiltonian $\hH(t) = \hH(t+2\pi/\wdr)$ can be recast into a numerically solvable eigenproblem for the so-called Floquet Hamiltonian \cite{sambe_1973}
\begin{align}
    \left[\hH(t)-i\partial_t \right]\ket{\phi_{\alpha}(t)}=\epsilon_{\alpha}\ket{\phi_{\alpha}(t)}.
    \label{Eq:FE}
\end{align}
The eigenvalues are the \textit{quasienergies} $\epsilon_\alpha$, and whose eigenvectors are the Floquet modes which are periodic functions of time with $\ket{\phi_{\alpha}(t)} = \ket{\phi_{\alpha}(t + 2\pi/\wdr)}$. In terms of these, the solution to the time-dependent Schr\"odinger is $\ket{\psi_\alpha(t)} = e^{-i\epsilon_\alpha t} \ket{\phi_\alpha(t)}$. Importantly, the solutions to \cref{Eq:FE} are only defined up to an integer multiple $k$ of the drive frequency $\wdr$, for if $\{\epsilon_\alpha, \ket{\phi_\alpha(t)}\}$ is a solution, then so is $\{\epsilon_{\alpha k} \equiv \epsilon_\alpha + k \wdr , \ket{\phi_{\alpha k}(t)} = e^{-i \wdr t} \ket{\phi_\alpha(t)}\}$, which is a consequence of the periodicity of the Floquet modes.

Information about the monochromatically driven system can be obtained from the quasienergy spectra. For example, two-tone spectroscopy experiments where a weak tone is used to probe the spectra of the driven system can be modeled in the linear response regime \cite{verney_et_al_2019}. In such experiments, probe-tone-induced transitions occur at frequency differences
\begin{align}
    \Delta_{\alpha\beta k} = \epsilon_\alpha-\epsilon_\beta + k \omega_d,
    \label{Eq:Dabk}
\end{align}  
provided that the operator corresponding to the probe tone, denoted generically as $\hat{X}$, has a nonzero matrix element between the corresponding Floquet modes. With the above notation, the corresponding matrix elements read
\begin{align}
X_{\alpha\beta k} = \frac{1}{T}\int_0^Tdt e^{-i(\epsilon_\alpha-\epsilon_\beta+k \omega_d)t}\langle\phi_\beta(t)|\hat X|\phi_\alpha(t)\rangle, \label{Eq:Xabk}
\end{align}
where $T=2\pi/\wdr$ is the period of the drive. This takes the form of a Fourier series coefficient $f_k = \frac{1}{T}\int_{0}^T dt' e^{-i k \frac{2\pi}{T} t'} f(t')$ of the matrix element of the operator $X$ between the two Floquet modes $\ket{\phi_{\alpha,\beta}(t)}$.

Numerically, the Floquet spectrum is efficiently obtained from the time-evolution operator over one period of the drive, which has a compact expression in terms of the Floquet modes \cite{grifoni_haenggi_1998}
\begin{align}
  \begin{split}
  \hat{U}(t+T,t) &= \mathcal{T} e^{-i \int_t^{t+T}  \hH(t') dt'} \\
  &= \sum_\alpha e^{-i \epsilon_\alpha T} \ket{\phi_\alpha(t)}\bra{\phi_\alpha(t)},
  \end{split}
\end{align}
where $\mathcal{T}$ is the time-ordering operator. According to the above expression, the Floquet modes at time $t=0$, $\ket{\phi_\alpha(0)}$, are the eigenvectors of $U(T,0)$, whereas the quasienergies are obtained modulo an integer multiple of $\wdr$ from the eigenvalues. The time-dependence over one period of the drive is obtained by propagating each mode $\ket{\phi_\alpha(0)}$ with the time-evolution operator $\hat{U}(t,0)$ in the interval $0 < t \leq T$. 

\begin{figure}[t]
  \includegraphics[width=\linewidth]{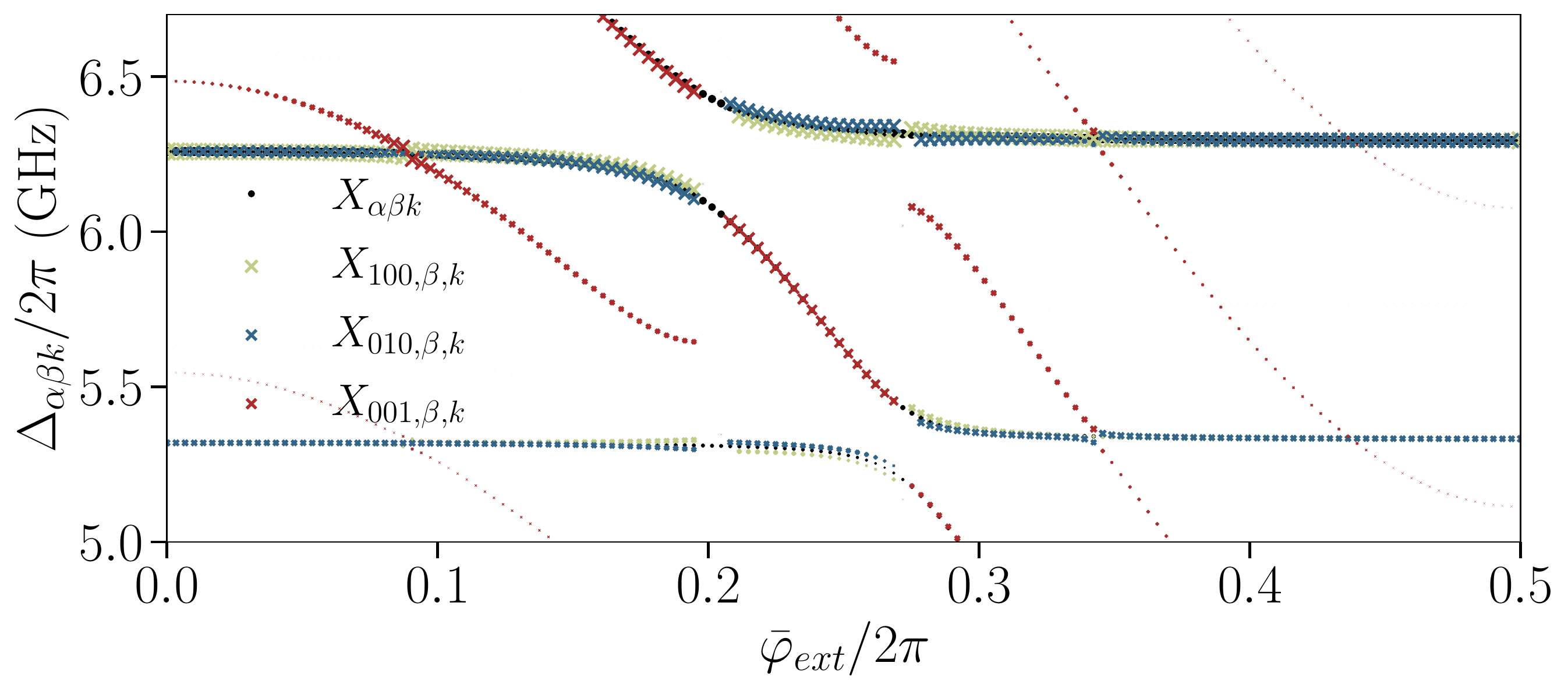}
  \caption{Floquet eigenspectrum for a rotating-wave approximation in which all photon-number nonconserving terms have been removed from the full circuit Hamiltonian analyzed in \cref{Sec:FullNum}. This figure is to be compared to the analogous result for the full Hamiltonian in \cref{fig:spectro_drive_non-RWA}.}
  \label{fig:spectro_drive_non-RWA-RWA-comparison}
  \end{figure}

To summarize, the steady-state dynamics can be obtained from the propagator $\hat{U}(t,0)$ over a single period of the drive, which makes the Floquet method  
an efficient alternative to numerical simulation of the dynamics over the complete gate time. Indeed, the period of the drive, on the order of $1\,\text{ns}$ is between two to three orders of magnitude shorter than the typical gate times. In this work we obtain the quantities above by using the QuTip implementation of the Floquet formalism \cite{JOHANSSON20131234}, to which we have contributed \cite{clc_github}, amended by a numerically efficient evaluation of the time-evolution operator developed by \textcite{shillito2020fast}. 

\section{Non-RWA effects in Floquet simulations of the full device}
\label{Ap:NonRWAEffects}
In this appendix we briefly discuss the role of counterrotating terms in the Floquet simulations of the full device Hamiltonian. Counterrotating terms (among which the parity-breaking cubic terms play an important role) in the coupler Hamiltonian induce an important correction to the coupler frequency, as can be seen by comparing \cref{fig:spectro_drive_non-RWA-RWA-comparison} to \cref{fig:spectro_drive_non-RWA}. This indicates, among other things, that a mere approximation of the coupler Hamiltonian as a Kerr nonlinear oscillator, as was done in the case of the toy model, would be insufficient for precise comparisons with experimental data. Moreover, the speedup obtained by using the Floquet method, together with the numerically efficient method for computing the time-evolution operator, enables us to study non-RWA effects efficiently as compared to full time-dynamics.

\bibliographystyle{apsrev4-1}
\bibliography{circuit_qed}
\end{document}